\documentclass[twocolumn]{aastex701}
\shorttitle{The origins of the UV emission of LRDs}
\shortauthors{Ando et al.}
\begin{document}

\title{The UV Side of Little Red Dots: Red, Compact, and Iron-Enhanced Rest-UV Emission with a Strong Downturn around Ly$\alpha$}

\author[orcid=0000-0002-4225-4477, gname=Makoto, sname=Ando]{Makoto Ando}
\affiliation{Institute for Cosmic Ray Research, The University of Tokyo, 5-1-5 Kashiwanoha, Kashiwa, Chiba 277-8582, Japan}
\email[show]{makoto.ando.astro@gmail.com}
%\correspondingauthor

%\author[orcid=0000-0000-0000-0001, gname=Yuichi, sname=Harikane]{Yuichi Harikane}
\author[0000-0002-6047-430X, gname=Yuichi, sname=Harikane]{Yuichi Harikane}
\affiliation{Institute for Cosmic Ray Research, The University of Tokyo, 5-1-5 Kashiwanoha, Kashiwa, Chiba 277-8582, Japan}
\email{hari@icrr.u-tokyo.ac.jp}

\author[orcid=0000-0003-1561-3814, gname=Harley, sname=Katz]{Harley Katz}
\affiliation{Department of Astronomy \& Astrophysics, University of Chicago, 5640 S Ellis Avenue, Chicago, IL 60637, USA}
\affiliation{Kavli Institute for Cosmological Physics, University of Chicago, Chicago IL 60637, USA}
\email{harleykatz@uchicago.edu}

\author[orcid=0000-0001-9840-4959,sname='Inayoshi']{Kohei Inayoshi}
\affiliation{Kavli Institute for Astronomy and Astrophysics, Peking University, Beijing 100871, China}
\email{inayoshi@pku.edu.cn}  

\author[0009-0003-4742-7060]{Takumi S. Tanaka}
\affiliation{Kavli Institute for the Physics and Mathematics of the Universe (WPI), The University of Tokyo Institutes for Advanced Study, The University of Tokyo, Kashiwa, Chiba 277-8583, Japan}
\affiliation{Department of Astronomy, Graduate School of Science, The University of Tokyo, 7-3-1 Hongo, Bunkyo-ku, Tokyo 113-0033, Japan}
\affiliation{Center for Data-Driven Discovery, Kavli IPMU (WPI), UTIAS, The University of Tokyo, Kashiwa, Chiba 277-8583, Japan}
\email{takumi.tanaka@ipmu.jp}


\begin{abstract}
Little Red Dots (LRDs) are candidates for growing supermassive black holes newly discovered by the James Webb Space Telescope (JWST), characterized by compact rest-optical morphology, V-shaped spectra, and broad Hydrogen Balmer lines.
While recently proposed BH-star/envelope models have made progress in explaining their optical features, their rest-UV emission, which is considered to originate from host galaxies, remains poorly investigated.
In this paper, we present a comprehensive analysis of the UV emission, including continuum shapes, emission line strengths, and morphology, using $\sim100$ LRDs selected from the JWST spectral archive.
Compared to star-forming galaxies at the same redshifts and UV magnitudes, LRDs show systematically redder UV slopes and more compact UV sizes, indicating that their UV emission cannot be explained solely by normal star-forming galaxies and requires a significant contribution from central red and compact emission.
From stacked spectra, we find that the Balmer break strength, UV slope, downturn depth around Ly$\alpha$, and Fe\,{\sc ii} equivalent width are positively correlated, while the UV size is anticorrelated with the Balmer break strength, suggesting that diversity in the UV continuum shape reflects the varying dominance of the central emission relative to its host.
We also measure $\mathrm{Fe\,\text{{\sc ii}}/Mg\,\text{{\sc ii}}}\sim8\text{--}10$, higher than in quasars at similar redshifts, further supporting a substantial contribution from the central component.
Spectral modeling suggests that the observed red UV continuum cannot be reproduced by host galaxy emission alone, but requires an additional very red continuum source ($\beta_\mathrm{UV}\sim0$), possibly nebular continuum emission leaking from dense ionized gas through a clumpy or porous neutral gas envelope.
\end{abstract}



\section{Introduction}
\label{sec:introduction}

% Overview of LRD
Little Red Dots (LRDs) are a new population of optically red and compact sources discovered by the James Webb Space Telescope (JWST) at $z>4$ (e.g., \citealt{Labbe2023,Kocevski2023,Harikane2023,Matthee2024,Greene2024}). Their red rest-frame optical colors (e.g., $\mathrm{F277W}-\mathrm{F444W}>1\,\mathrm{mag}$; \citealp{Greene2024}), broad Balmer lines, and unresolved morphology even in JWST images suggest that they are growing supermassive black holes (SMBHs) in the early universe \citep{Inayoshi2025c}. Intensive explorations of LRDs have revealed that they are surprisingly abundant ($10^{-(4\text{--}5)}\,\mathrm{Mpc^{-3}}$ at $z\sim4\text{--}6$; \citealp{Kokorev2024,Kocevski2025,Umeda2026}) and exist across a wide range of cosmic epochs from the present day ($z\sim0.1$; \citealp{Lin2026,X.Ji2026}) to the cosmic dawn, even at $z\sim10$ \citep{Taylor2025b,Tanaka2025}, making LRDs a key population for understanding the formation of SMBHs through cosmic history. 

The nature of LRDs remains unclear because they are faint across most wavelengths except for the rest-frame ultraviolet (UV) to optical range. Unlike other AGN populations, they are not detected in X-rays from a hot corona around the accretion disk \citep{Ananna2024,Yue2024,Akins2025,Kokubo2025,Maiolino2025} as well as MIR emission from a dust torus \citep{Akins2025,Xiao2025,Setton2025a,Casey2025}. The X-ray non-detection, even with ultradeep Chandra observation and stacking analysis, implies the extremely high obscuration due to Compton thick gas ($N_\mathrm{H}>10^{24}\,\mathrm{cm^{-2}}$; \citealp{Sacchi2025}). The majority of them also lack rest-frame hundred-day-scale time variability \citep{Kokubo2025,Liu2026,Z.Zhang2025a}, although possible variability over rest-frame several-to-tens of years has been reported \citep{Furtak2025,Z.Zhang2025b,X.Ji2025}. These unusual properties may suggest that LRDs are a distinct population from known AGNs.

% Optical side of LRD
The rest-frame UV-to-optical spectral energy distribution (SED) of LRDs is characterized by V-shaped continua with red optical slope and blue UV slope (e.g., \citealp{Kokorev2024,Kocevski2025,Hainline2025}). The optical spectrum shows a strong Balmer break \citep{Setton2025b}, attributed to a red optical slope.
The Balmer break strengths for some extreme LRDs are twice as strong as those of massive quiescent galaxies and cannot be reproduced by any stellar population \citep{Naidu2025, de_Graaff2025a}. Another characteristic feature in the rest-optical is the presence of absorption features in broad Balmer lines (e.g., \citealp{Matthee2024,Matthee2026,Kocevski2025}). 

%BH*/BH envelope model
Recently, Black Hole Star (BH*) or BH envelope, a group of empirical models that self-consistently explain these observational features of the LRD spectrum, has been proposed \citep{Naidu2025,Inayoshi2025,Inayoshi2026,Kido2025}. A BH envelope model assumes that an accreting BH is surrounded by a dense (e.g., $N_\mathrm{H}\sim10^{24\text{--}26}\,\mathrm{cm^{-2}}$), dust-free, and turbulent ($\sim 500\,\mathrm{km\,s^{-1}}$) gas. These assumptions are consistent with the absence of X-ray and dust emission and the presence of Balmer absorption lines. In this model, a strong Balmer break feature is explained by the intrinsic shape of blackbody radiation from the pseudo-photosphere at $\sim5000\,\mathrm{K}$ with the Balmer absorptions \citep{Naidu2025,de_Graaff2025c}.
% or, alternatively, the Paschen and Brackett continuum emitted from low-temperature ($<10000\,\mathrm{K}$) gas \citep{Seneppen2026}. 
In addition, Balmer lines may be significantly broadened due to Thomson scattering effect (e.g., \citealp{Rusakov2026,Chang2026}) within a dense ionized gas ($n_\mathrm{e}\gtrsim10^{8}$) near the BH accretion disk, mitigating the over-abundance of massive BHs in the early Universe (cf. \citealp{Perez-Gonzalez2024,Ma2025,Ma2026}).

%UV emission from LRD
While the BH envelope model well reproduces the characteristic spectra of the optical SED of LRDs, the origins of UV emission remain rather unconstrained. In line with the BH envelope model, accounting for its faintness, UV emission is thought to originate from the low-mass ($\log(M{*}/M_{\odot})\sim 8\text{--}9$; \citealp{Y.Zhang2025}) host galaxy, such as star-forming dwarfs \citep{Naidu2025,Sun2026} or a nuclear star cluster around the neutral gas envelope \citep{Inayoshi2026}. This assumption is consistent with observations showing the extended emission beyond the point spread function (PSF) in the rest-UV images \citep{Labbe2024,Y.Zhang2025,Cloonan2026,Torralba2026a}. On the other hand, from morphological analysis of LRDs, it has been reported that the emission from unresolved components accounts for $20\text{--}40\%$ of the far-UV flux \citep{Y.Zhang2025,Cloonan2026}. This indicates that some fraction of UV emission is leaked from the dense neutral gas envelope. Such a leaked emission may be accounted for in the envelope model, instead of a fully covered geometry, by assuming the envelope has holes or is clumpy \citep{Tang2026,Ji2026}.

Several key limitations, however, still remain in previous studies of LRDs' UV emission.
First, some studies have been based on individual objects or small samples, preventing us from examining whether proposed LRD structures can explain not only the UV properties of individual objects but also the diversity of LRD populations as a whole. Second, many statistical studies rely solely on photometric data, which introduces uncertainties in LRD selection and lacks spectroscopic information necessary to infer the detailed physical state of LRDs. Third, different observational diagnostics, such as UV-to-optical continua, emission lines, and morphology, have often been studied separately. As a result, the origin of the UV emission, including the host and central source, has not yet been consistently constrained in a unified framework. 

In this study, we aim to address these issues through a comprehensive analysis focusing on the rest-frame UV properties of LRDs. We use LRD samples selected from JWST archival spectra presented by \citet{de_Graaff2025c}. Our sample consists of $\sim100$ LRDs, sufficiently large to discuss UV properties of LRDs at a population level. We first measure UV slopes and sizes consistently as in the literature on high-redshift galaxies, and clarify that the LRD hosts are a distinct population from normal star-forming galaxies.
Then, we perform spectral stacking of LRDs to examine emission-line properties and detailed continuum shapes. Furthermore, we integrate our findings and propose SED models that explain the observed UV spectra, advancing our understanding of the internal structure of LRDs in line with the BH envelope scenario.

The structure of this paper is as follows. In Section~2, we describe the imaging/spectroscopic data and LRD sample selection used in this paper. In Section~3, we explain how to calculate the UV slopes and the UV sizes of the LRDs. We also present the methodology for spectral stacking. The results of these analyses are shown in Section~4. We discuss the origins of UV emission of LRDs in Section~5, and Section~6 is devoted to a summary and conclusions. Throughout this paper, we assume a flat $\Lambda$ cold dark matter cosmology with ($\Omega_\mathrm{m}$, h) = (0.3, 0.7) and use AB magnitudes \citep{Oke1983}.

\begin{figure*}[t!]
\centering
\includegraphics[width=2\columnwidth]{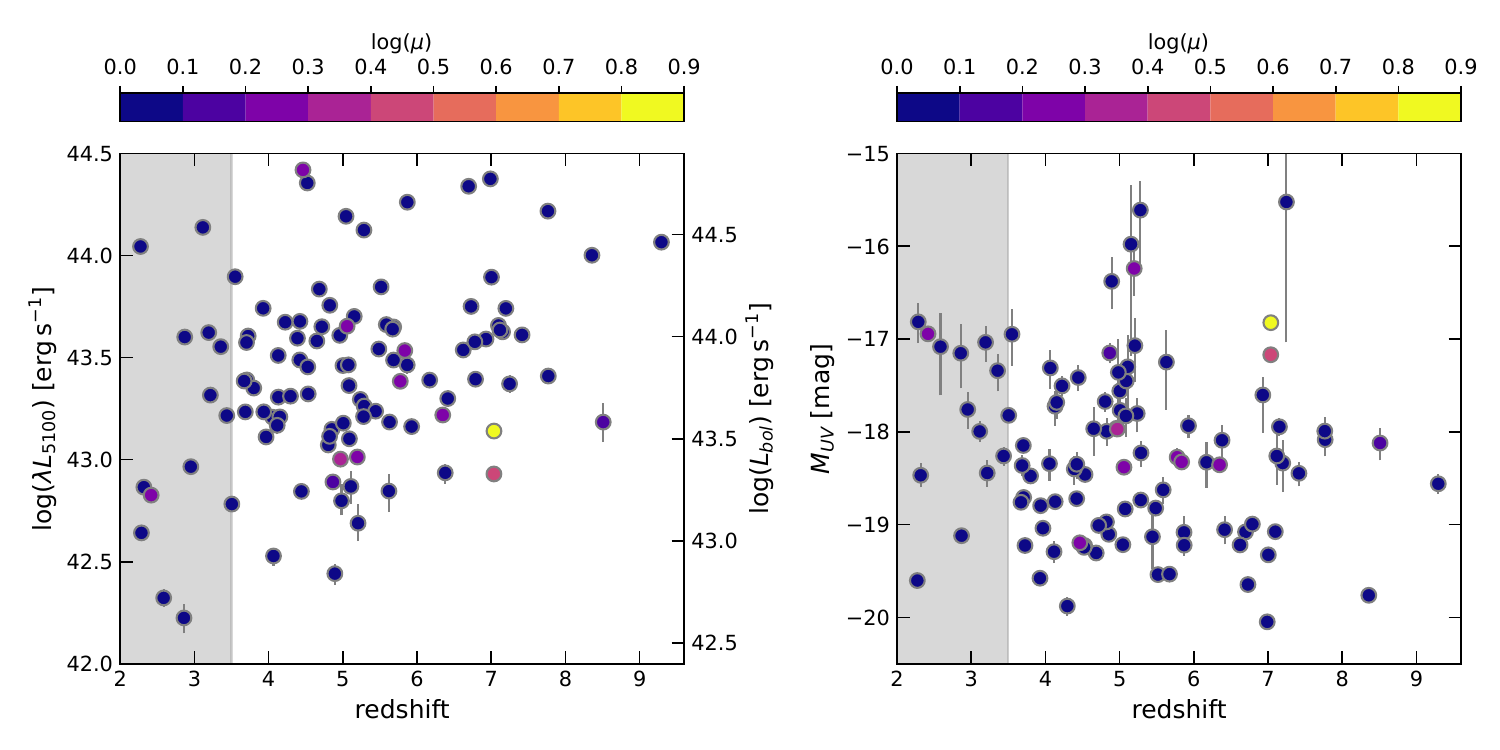}
\caption{Overview of the LRD sample used in this paper. The LRDs are basically selected in terms of the V-shape spectra and compact morphology in the F444W images by \citet{de_Graaff2025c}. \textit{Left panel:} optical luminosities at $5100\,\mathrm{\AA}$ against redshifts color-coded by magnification factors.
The second y-axis shows the corresponding bolometric luminosities converted with the bolometric correction of $L_\mathrm{bol}/L_{5100}=2.5$, calculated from the results in \citet{de_Graaff2025c}.
We remove the lower-redshift ($z<3.5$, highlighted in gray) sample from the main analysis to investigate the rest-frame UV features of LRDs. \textit{Right panel:} same as the left panel, but for the UV absolute magnitudes estimated from NIRSpec PRISM spectra (see Section~\ref{sec:analysis_uvslope}).}
\label{fig:sample}
\end{figure*}

\section{Sample and dataset}
\label{sec:sample}
We use an LRD sample presented by \citet{de_Graaff2025c}. They select 116 LRDs from NIRSpec MSA spectra in version 4.4 of the Dawn JWST Archive (DJA; \citealp{Brammer2025_DJA}). We first describe the spectroscopic and imaging data used in \citet{de_Graaff2025c} and in this paper in Sections~\ref{sec:data_spec} and \ref{sec:data_image}, respectively. Then, our LRD samples are described in Section~\ref{sec:data_lrd}.

\subsection{Spectroscopic data}
\label{sec:data_spec}

We use version 4.4 of the spectroscopic compilation catalog of DJA. All spectra are reduced with \texttt{msaexp} following the methods described in \citet{Heintz2025} and \citet{de_Graaff2025b}. We use NIRSpec PRISM/CLEAR spectra. A customized reference file for the flux calibration enables spectral extraction beyond the nominal wavelength range (e.g., $\sim 5.5\,\mathrm{\mu m}$, \citealp{Valentino2025}). The 1D spectra are extracted with a wavelength-dependent slit loss correction based on the extraction kernel together with the information of the source position in the shutter, which performs well at least for point sources (e.g. \citealp{de_Graaff2025a}).

\subsection{Imaging data} 
\label{sec:data_image}

For imaging data, we mainly use JWST/NIRCam image mosaics from the DJA version 7. All DJA images have been reduced with \texttt{grizli} \citep{Brammer2008} following the methodology described in \citet{Valentino2023}. NIRCam images are collected from various imaging programs: CANUCS (Sarrouh et al. 2025), CEERS (Finkelstein et al. 2025), JADES (Eisenstein et al. 2023), NEXUS (Shen et al. 2024), PRIMER (GO-1837; PI: Dunlop), and UNCOVER (Bezanson et al. 2024), as well as pure parallel programs such as PANORAMIC (Williams et al. 2024a). For relatively large fields (i.e., CEERS and PRIMER), image pixel scales are $0.04''/\mathrm{pix}$. For other fields, pixel scales of $0.02''/\mathrm{pix}$ and $0.04''/\mathrm{pix}$ are applied to images in the short wavelength (SW; $<2.4\,\mathrm{\mu m}$) and the long wavelength (LW; $>2.4\,\mathrm{\mu m}$) channels, respectively. 

\subsection{LRD sample} 
\label{sec:data_lrd}

%de Graaf+25 and selection
In this paper, we use the spectroscopically selected LRD sample in \citet{de_Graaff2025c}. They select LRDs based on two criteria: the V-shaped continuum and compactness. First, to select objects with V-shaped continua, they perform power-law fitting at rest-frame $1200\,\mathrm{\AA}<\lambda<7000\,\mathrm{\AA}$ for all available DJA NIRSpec/PRISM spectra. They adopt a broken power-law model whose two slopes, $\beta_\mathrm{UV}$ and $\beta_\mathrm{opt}$, are connected at the Balmer limit. They mask strong emission lines and sample these two slopes using the Monte Carlo Markov Chain (MCMC) procedure. The V-shaped spectrum is defined as those whose the 95 percent MCMC samplings satisfy: (i) the optical slope is red ($\beta_\mathrm{opt}>0$),  (ii) the UV slope is blue ($\beta_\mathrm{UV}<-0.2$), and (iii) the optical side is sufficiently red to UV side ($\beta_\mathrm{UV}-\beta_\mathrm{opt}>0.5$). For 247 selected spectra in this step, they require compactness criteria in the F444W image: (i) the flux ratio measured with $0.2''$ and $0.1''$ apertures is small ($f_\mathrm{F444W}(0.2'')/f_\mathrm{F444W}(0.1'')<1.7$), or (ii) the two-component 2D-light profile fitting with a point source and a S\'ersic model suggests more than 50\% flux in F444W is dominated by point source component. The latter is examined only in the five CANDELS and Abell-2744 fields, where the immediate PSF model \citep{Weibel2024} is available. After screening for duplicated objects and performing a visual inspection, 116 LRDs are selected.

In Figure~\ref{fig:sample}, we show the optical luminosity and UV absolute magnitude at rest-frame $\lambda=1500\,\mathrm{\AA}$ of the selected LRDs. In this paper, we use LRDs only at $z>3.5$ to capture the rest-frame UV component within PRISM's wavelength coverage, although the original catalog includes them down to $z=2$. We note that we correct for the gravitational magnification factor $\mu$ using values reported in \citet{de_Graaff2025c}.

\section{Analysis}
\label{sec:analysis}

\subsection{UV slope measurement}
\label{sec:analysis_uvslope}

We measure the UV slope $\beta_\mathrm{UV}$ and $M_\mathrm{UV}$ using PRISM/CLEAR spectra at rest-frame $1340\text{--}2700\,\mathrm{\AA}$. This wavelength range is comparable to those used in the literature to estimate the UV slope of normal star-forming galaxies (e.g., \citealp{Bouwens2014, Saxena2024}), enabling us to make a fair comparison. Following \citet{Saxena2024}, we mask wavelength ranges which might be contaminated by strong emission lines: $1440\text{--}1590\,\mathrm{\AA}$, $1620\text{--}1680\,\mathrm{\AA}$, and $1860\text{--}1980\,\mathrm{\AA}$.

We adopt a simple power-law model:
\begin{equation}
    f_\mathrm{\lambda,rest}=f_\mathrm{\lambda,0}\left( \frac{\lambda}{1500\,\mathrm{\AA}}\right) ^{\beta_\mathrm{UV}},
\end{equation}
where $f_\mathrm{\lambda,0}$ is the normalization at the pivot wavelength of $1500\,\mathrm{\AA}$. We perform MCMC fitting and derive the median and 68\% credible intervals of the posterior distribution. The UV absolute magnitudes are estimated from $f_\mathrm{\lambda,0}$.

We remove three LRDs whose UV spectra are largely affected by a detector gap. We note that the wavelength range adopted here is not the same as that in \citet{de_Graaff2025c}, who include the longer wavelength up to $\sim3650\,\mathrm{\AA}$, although the estimated UV slopes are consistent within uncertainties. We also test whether the UV slope estimation is significantly affected by additionally masking the UV Iron lines ($2200\text{--}3100\,\mathrm{\AA}$), as presented in Section~\ref{sec:analysis_stack}, the results remain unchanged at the population level within the uncertainties.

\begin{figure*}[ht!]
\centering
\includegraphics[width=2.1\columnwidth]{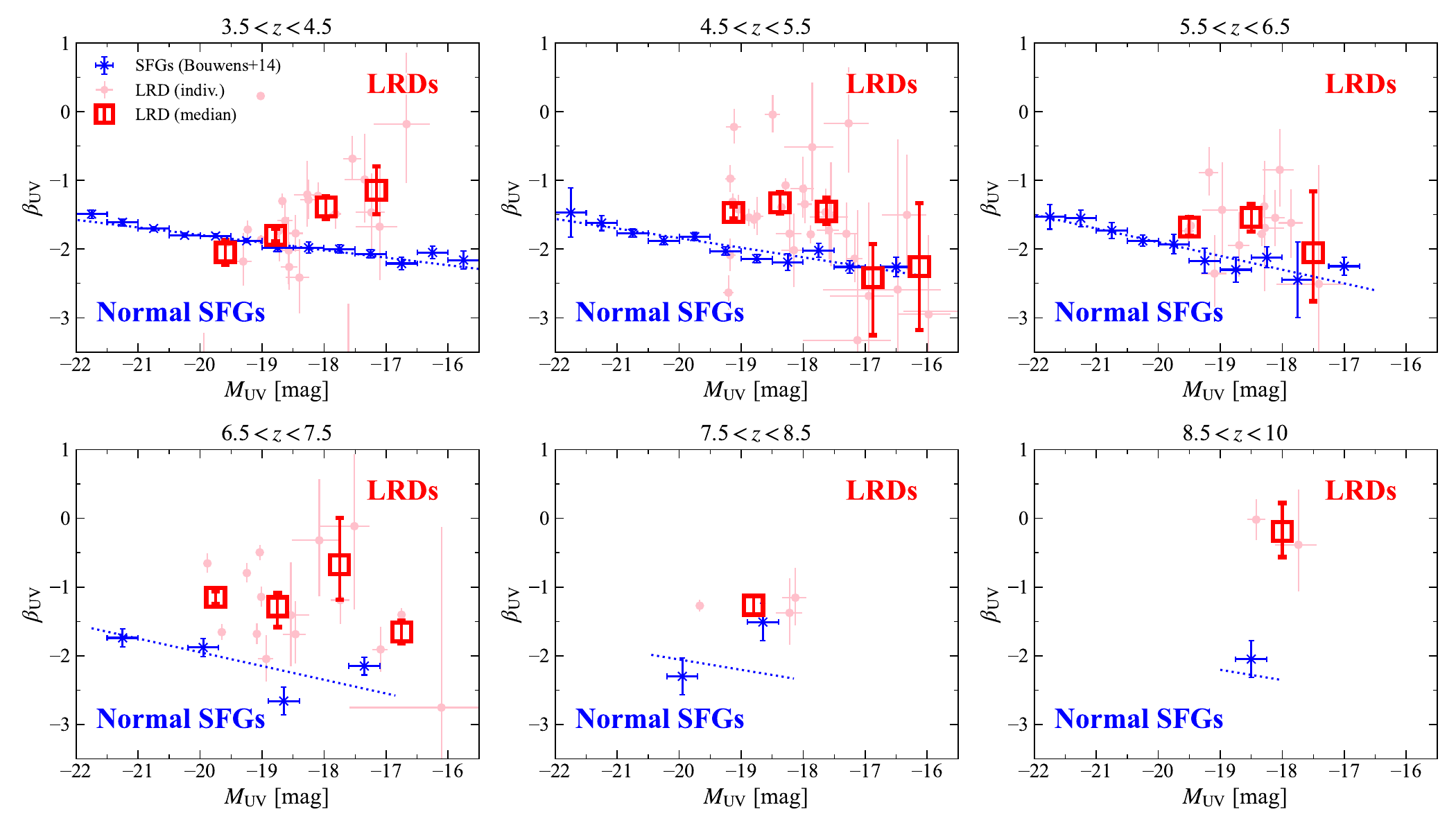}
\caption{Relation between UV slope $\beta_\mathrm{UV}$ and the UV magnitude at $1500\,\mathrm{\AA}$ for six redshift bins. The red dots are individual LRDs, and the red open squares are their medians in given magnitude bins. The medians are shown only when more than one data point falls into a magnitude bin. For comparison to normal star-forming galaxies (SFGs), we also show the median $\beta_\mathrm{UV}$-$M_\mathrm{UV}$ relations for Lyman break galaxies selected by \citet{Bouwens2014} as blue crosses and dotted lines. Since \citet{Bouwens2014} lacks the relation for the $8.5<z<10$ bin, we instead show it for $7.5<z<8.5$. As a general trend, UV slopes of LRDs are redder than those of normal SFGs regardless of redshifts.}
\label{fig:uvslope}
\end{figure*}

\subsection{UV size measurement}
\label{sec:analysis_size}

We aim to measure the rest-frame UV sizes of the LRDs by 2D light-profile fitting and compare them to those of normal star-forming galaxies. Some studies have reported that a fraction of LRDs show blue companions spatially offset to a central point-like component, whose physical origins are unclear \citep{Golubchik2025,Rinaldi2025,C.Chen2025,Baggen2026}. Here, we examine the UV emission associated with the central component, as the MSA spectra primarily capture it rather than offset companions.

Following \citet{Shibuya2015}, we use the shortest wavelength images available that fully covered by the rest-frame $1500\text{--}3000\,\mathrm{\AA}$, namely, F090W ($3.5<z<4.3$), F115W ($4.3<z<5.8$), F150W ($5.8<z<7.9$), F200W ($z>7.9$). The signal-to-noise ratio (SN) of each LRD in the UV image is measured with the $0.2''$ aperture. For each LRD, we make UV and F444W cutout images centered on its position reported by \citet{de_Graaff2025c} with a side length of $2.4''$. The PSF in each filter is created from the median-stacked image of natural stars. The typical full-width at half-maximum (FWHM) for the rest-frame UV images is $0.05''\text{--}0.08''$, while that for the F444W image is $0.16''$.

We perform single S\'ersic profile fitting to the rest-frame UV images using \texttt{Galight} \citep{Ding2021}. \texttt{Galight} automatically detects sources in a given image cutout and simultaneously fits the S\'ersic profile to each detected source, which is useful for separating the central component from other companions. To pinpoint the UV emission associated with the central component, we first run object detection on both the UV and F444W cutouts\footnote{For object detection, we require at least five contiguous pixels with $\mathrm{SN}>3$.}. Then, we treat the nearest UV source within $0.12''$ (i.e., slightly smaller than the PSF size of the F444W image) from each LRD position in F444W as that associated with the LRD. If no UV source is detected due to a low SN, we assume the UV source is at the same position as the counterpart in the F444W image.

We model the 2D light profiles of all detected objects in UV images with the S\'ersic profile \citep{Sersic1963}:
\begin{equation}
    I(r)=I_{e}\exp \left( -b_{n}\left[ \left\{\left(\frac{r}{r_{e}}\right)^{1/n}-1 \right\} \right]  \right),
\end{equation}
where $I_{e}$ is the surface brightness at the half-light radius $r_{e}$, and $n$ is the S\'ersic index. $b_{n}$ is a coefficient that makes $r_{e}$ enclose a half of the total flux.

Model fitting is performed by an MCMC procedure with the following parameters: source central position, total flux, half-light radius along the major axis, S\'ersic index, and ellipticities. During fitting process, we require $r_{e}$, $n$, and ellipticity $q$ to be $0.01^{''}<r_{e}<2^{''}$, $0.01<n<4$, and $q>0.5$, respectively. We take the mode and 68\% highest probability density interval from the posterior distribution. If the mode of $r_{e}$ converges to the lower limit of the fitting range, we regard the source as unresolved and instead report a $1\sigma$ upper limit. The derived half-light radius along the major axis is converted to the circularized one in the physical scale using ellipticities. We correct for the gravitational lensing effect by dividing apparent sizes by $\sqrt{\mu}$. We note that 10 LRDs with unsuccessful fits due to low SN or bright neighbors are removed by visual inspection.

\subsection{Spectral stacking}
\label{sec:analysis_stack}

The UV emission and absorption lines provide hints at understanding the properties of the primary UV source. Since the UV light from LRDs is usually faint compared to optical light, we stack PRISM spectra to examine the UV spectral shape beyond the slope and detect emission lines that provide clues to understanding the origins of their UV emission.

We convert all PRISM spectra into rest-frame and re-bin them to $\Delta \lambda=5\,\mathrm{\AA}$ (cf. \citealp{Perez-Gonzalez2026}). After normalizing the spectra at rest-frame $3000\,\mathrm{\AA}$, we perform median stacking. The uncertainties of the stacked spectrum are estimated by bootstrapping. %We also make stacked spectra of LRDs with blue ($\beta_\mathrm{UV}<-1.5$) and ($\beta_\mathrm{UV}>-1.5$) UV slopes.

To characterize the UV continuum of stacked spectra and measure the equivalent widths (EWs) of emission lines such as C\,{\sc iii}] and C\,{\sc iv}, we estimate the continuum shape assuming the power-law form. We adopt wavelength masks used in Section~\ref{sec:analysis_uvslope}. Since the detection of Fe\,{\sc ii} emission line complex has been often reported in LRD spectra (e.g., \citealp{Naidu2025,Perez-Gonzalez2026,Torralba2026b}), we additionally mask $2200\text{--}3100\,\mathrm{\AA}$ to avoid the UV slope from being biased to be redder. Instead, we extend the fitting range to $3500\,\mathrm{\AA}$ to increase SN. Then, following \citet{de_Graaff2025a,de_Graaff2025c}, we quantify the Balmer break strength of the stacked spectrum as the ratio of flux density $f_{\nu}$ between $3620\text{--}3720\,\mathrm{\AA}$ and $4000\text{--}4100\,\mathrm{\AA}$. We note that the Balmer break strength correlates with the UV-to-optical flux ratio, which is defined between $1500\,\mathrm{\AA}$ and $5100\,\mathrm{\AA}$.

To further examine whether UV properties correlate with continuum shapes, we make LRD subsamples based on individually measured UV slope and Balmer break strength. For UV slope, we divide the LRD sample into red- ($\beta_\mathrm{UV}>-1.5$) and blue- ($\beta_\mathrm{UV}<-1.5$) UV-slope subsamples. For Balmer break strength, we select strong- ($>2$), intermediate- ($1.2\text{--}2$), and weak- ($<1.2$) Balmer-break subsamples.

\section{Results}
\label{sec:result}

\begin{figure}[ht!]
\centering
\includegraphics[width=\columnwidth]{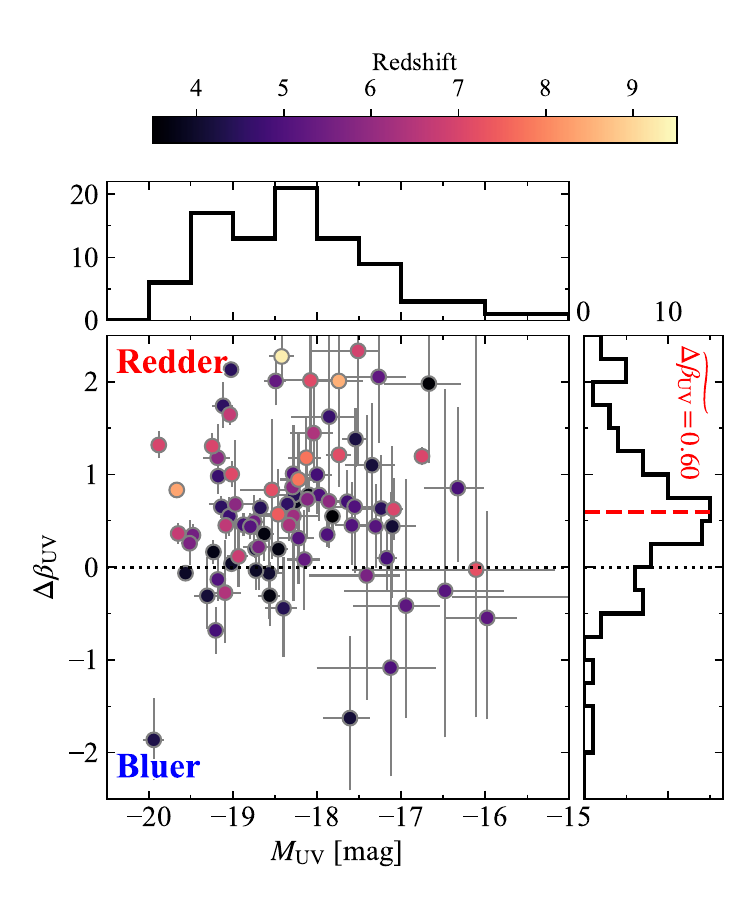}
\caption{Relative values of the UV slopes of the LRDs against those calculated from the $M_\mathrm{UV}$-$\beta_\mathrm{UV}$ relation of normal SFGs, $\Delta \beta_\mathrm{UV}=\beta_\mathrm{UV,\,LRD}-\beta_\mathrm{UV,\,base}$, color-coded by redshift. The histograms of $M_\mathrm{UV}$ and $\Delta \beta_\mathrm{UV}$ are shown in the top and right subplots, respectively. The black dotted and red dashed horizontal lines indicate zero and median $\Delta \beta_\mathrm{UV}$ values, respectively. A positive (negative) $\Delta \beta_\mathrm{UV}$ means that a given LRD has redder (bluer) UV slope compared to an average star-forming galaxy with a similar redshift and $M_\mathrm{UV}$. The $\Delta \beta_\mathrm{UV}$ distribution is clearly skewed towards the positive values with the median value of $0.60$, indicating that LRDs have redder UV slopes compared to normal star-forming galaxies at a population level.
}
\label{fig:delta-uvslope}
\end{figure}

\subsection{UV slope}
\label{sec:result_uvslope}

Figure~\ref{fig:uvslope} shows the relation between the measured UV slopes and UV absolute magnitude at $1500\,\mathrm{\AA}$ for six redshift bins.
The UV slopes of normal star-forming galaxies \citep{Bouwens2014} are $-2.5<\beta_\mathrm{UV}<-1.5$ with a decreasing trend with $M_\mathrm{UV}$ (i.e., bluer slopes for fainter sources), which is due to less evolved stellar populations, lower metallicity, and weaker dust reddening (e.g., \citealp{Gonzalez2012,Bouwens2012}). Compared to them, LRDs have redder UV slopes, $\beta_\mathrm{UV}\sim -1.5$, without clear trends with UV magnitude. One fourth of LRDs have very red UV slopes of $\beta_\mathrm{UV}>-1$, and some extremes reach $\beta_\mathrm{UV}\sim0$.

We further examine the relative redness of LRDs' UV slopes compared to normal star-forming galaxies. For each LRD, we define the baseline UV slope $\beta_\mathrm{UV,\, base}$ from the $\beta_\mathrm{UV}\text{--}M_\mathrm{UV}$ relation of star-forming galaxies at a given redshift bin (i.e., blue dashed lines in Figure~\ref{fig:uvslope}; \citealp{Bouwens2014}) and calculate $\Delta \beta_\mathrm{UV}=\beta_\mathrm{UV,\,LRD}-\beta_\mathrm{UV,\, base}$. In Figure~\ref{fig:delta-uvslope}, we show the distribution of $\Delta \beta_\mathrm{UV}$. As expected, $\Delta \beta_\mathrm{UV}$ is skewed toward positive values (i.e., red UV slopes) with the median $\Delta \beta_\mathrm{UV}$ of $0.60$, meaning that LRDs have redder UV slopes compared to the normal star-forming galaxies at a population level.

\begin{figure*}[ht!]
\centering
\includegraphics[width=2.1\columnwidth]{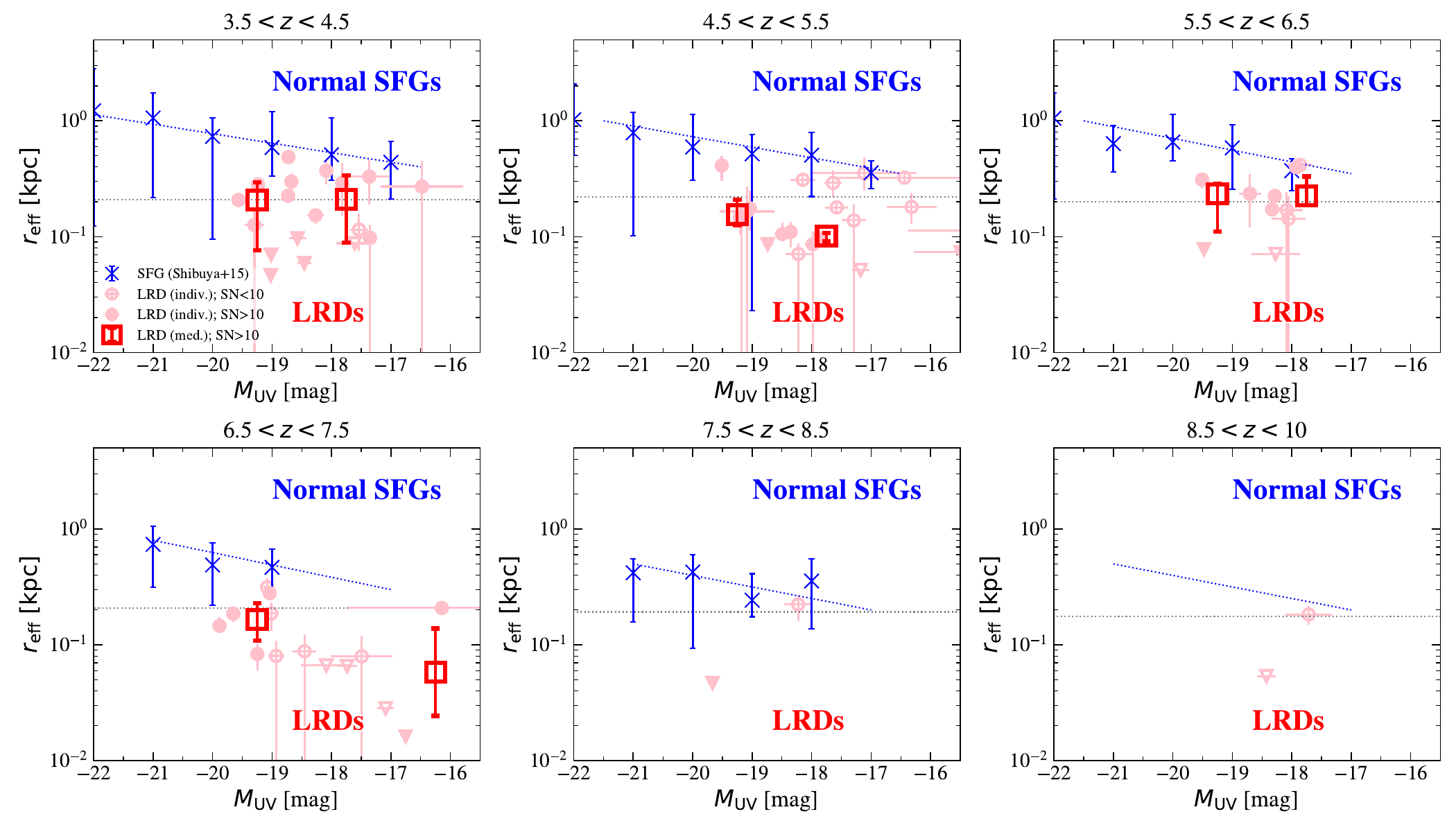}
\caption{Relation between effective radii $r_\mathrm{eff}$ and the UV magnitude at $1500\,\mathrm{\AA}$ for six redshift bins. The red-filled (open) dots are individual LRDs with $\mathrm{SN}>10$ ($\mathrm{SN}<10$) in the UV images, and the downward triangles indicate $1\sigma$ upper limits for unresolved sources. Red open squares are median sizes computed in log-space for the $\mathrm{SN}>10$ subsamples in given magnitude bins. 
We conservatively use $1\sigma$ upper limits for unresolved sources in the calculation; therefore, the true median values could be smaller. Medians are shown only when more than one data point falls into a magnitude bin. For comparison with normal SFGs, we also show the median size-luminosity relations presented in \citet{Shibuya2015} as blue crosses and dotted lines. Since \citet{Shibuya2015} lacks the relation for the highest redshift bin at $8.5<z<10$, we instead show it for $7.5<z<8.5$. Gray dotted lines are the typical sizes of the point spread function (PSF) for each redshift. The rest-UV sizes of LRDs are clearly smaller than normal SFGs.
}
\label{fig:size}
\end{figure*}

\subsection{UV Size}
\label{sec:result_size}

Figure~\ref{fig:size} shows the relation between the UV effective radius and UV magnitude.
We find that the typical UV sizes of LRDs are several hundred pc, which are comparable to or even smaller than the PSF.\footnote{A fraction of LRDs may be unresolved and therefore better described by a PSF than by a S\'ersic profile, which could lead to systematically overestimated sizes when applying simple profile fitting. While a more rigorous characterization would require detailed two-dimensional image decomposition and careful treatment of the PSF, the current size estimates would be sufficient to examine whether LRDs are systematically more compact than typical galaxies.}
They are much smaller than typical star-forming galaxies in \citet{Shibuya2015}, with similar UV magnitudes and redshifts. For instance, LRDs with $\mathrm{SN}>10$ are, on average, $\sim 0.5\,\mathrm{dex}$ smaller than normal star-forming galaxies.

\begin{figure}[ht!]
\centering
\includegraphics[width=\columnwidth]{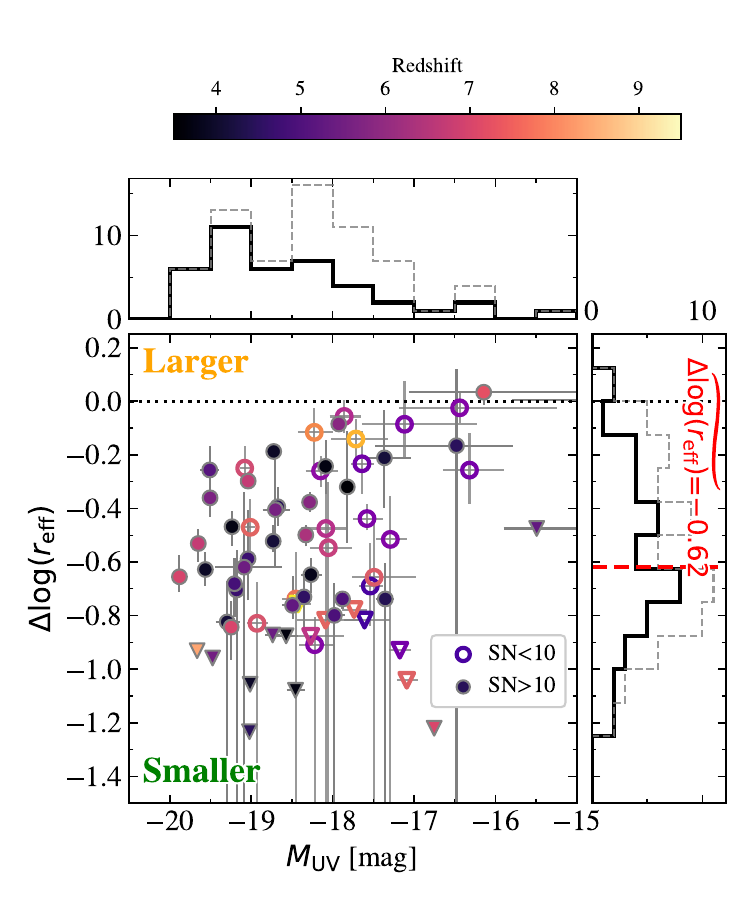}
\caption{Same as Figure~\ref{fig:delta-uvslope} but for LRDs' relative effective radii against those of normal SFGs. Taking the $M_\mathrm{UV}$-$r_\mathrm{eff}$ relation of normal SFGs as base lines, we calculate $\Delta \log(r_\mathrm{eff})=\log(r_\mathrm{eff}/r_\mathrm{eff,\,base})$. LRDs with high ($>10$) and low ($<10$) SN in UV images are shown by the filled and open marks, respectively, and their distributions along the two axes are indicated by black and gray on the subplots. For unresolved sources indicated by downward triangles, we show $1\sigma$ upper limits. The median $\Delta \log(r_\mathrm{eff})$ value for the $\mathrm{SN}>10$ subsample is shown by the red dashed line. A positive (negative) $\log(r_\mathrm{eff})$ means that a given LRD has a larger (smaller) UV size compared to an average SFG with a similar redshift and $M_\mathrm{UV}$. The $\Delta \log(r_\mathrm{eff})$ distribution is clearly skewed towards the negative values with the median of $-0.62$, indicating that LRDs are by several times smaller in UV than normal SFGs at a population level.
}
\label{fig:delta-size}
\end{figure}

In the same manner as $\beta_\mathrm{UV}$, we examine LRDs' relative effective radii against those of normal star-forming galaxies. Accounting for the $M_\mathrm{UV}$-$r_\mathrm{eff}$ relation of \citet{Shibuya2015} as baselines, we calculate $\Delta \log(r_\mathrm{eff})=\log(r_\mathrm{eff}/r_\mathrm{eff,\, base})$.
As shown in Figure~\ref{fig:delta-size}, the $\Delta \log(r_\mathrm{eff})$ distribution is clearly skewed towards the negative values (i.e., smaller sizes) with the median of $-0.62$, indicating that LRDs are several times smaller in UV than normal star-forming galaxies at a population level.

\subsection{Stacked Spectra}
\label{sec:result_stack}

\begin{figure*}[ht!]
\centering
\includegraphics[width=2.1\columnwidth]{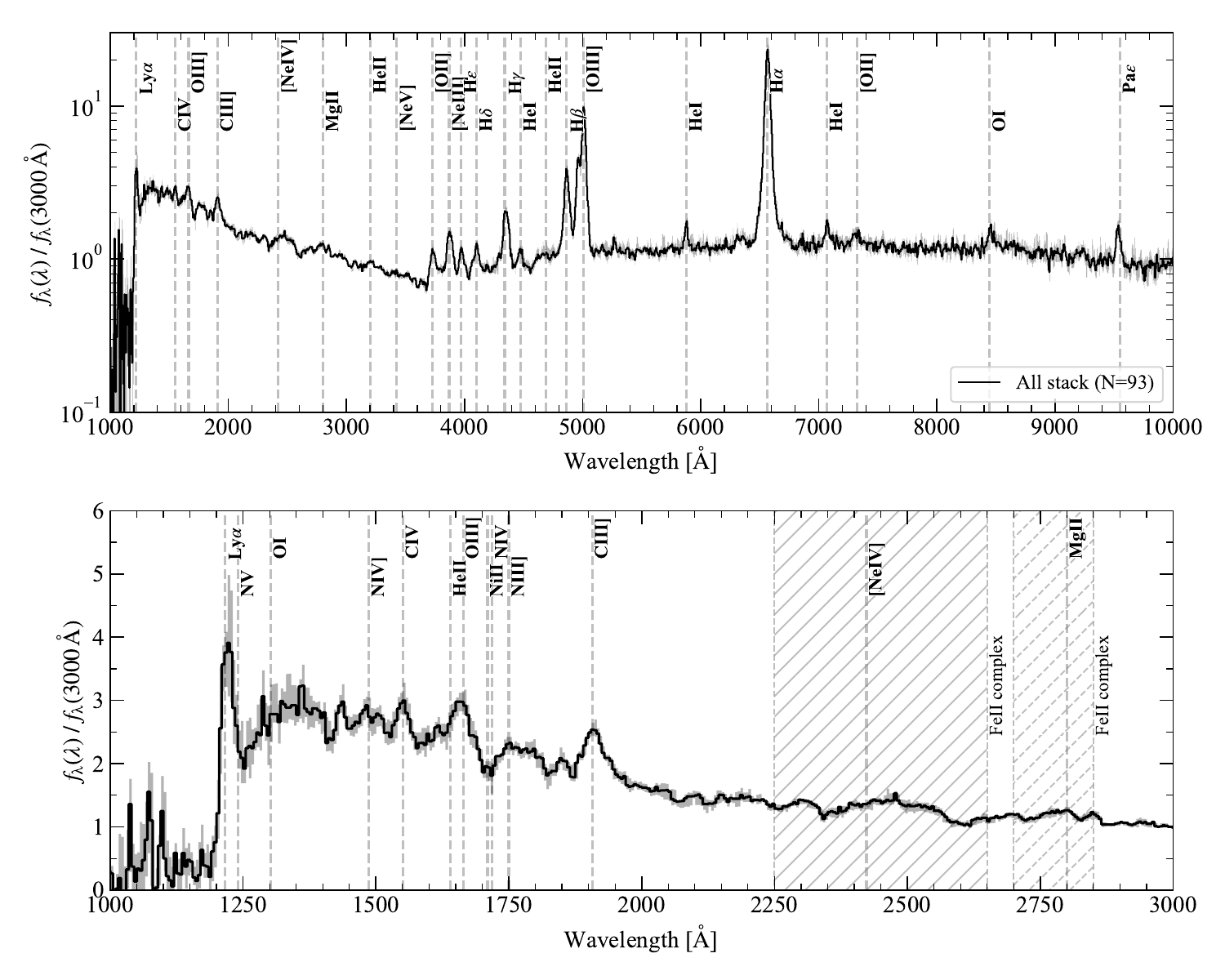}
\caption{Stacked spectrum of all LRDs at $z>3.5$ (N=93). The top panel shows the entire spectral range from UV to optical, while the bottom panel focuses on the UV range. We adopt the median stacking after normalizing individual spectra at $3000\,\mathrm{\AA}$. Uncertainties of the stacked spectrum estimated by bootstrap are shown by gray shades. Representative emission lines are shown by black dashed lines. The gray background hatch in the bottom panel indicates Fe\,{\sc ii} complexes.}
\label{fig:stack_all}
\end{figure*}

Figure~\ref{fig:stack_all} shows the stacked PRISM spectrum for all LRD samples at $z>3.5$ ($N=93$). Spectral stacking greatly improves SN and enables us to discuss emission lines and spectral shapes at the population level. At optical wavelengths, the stacked spectrum shows many emission lines, including the Balmer series and lines from helium, carbon, nitrogen, oxygen, and neon. The clear Balmer break, a representative feature of LRD spectra, is present. In the UV range, we find some emission lines such as Ly$\alpha$, C\,{\sc iii]}, C\,{\sc iv}, and O\,{\sc iii}] possibly blended with He\,{\sc ii}. At $2300\text{--}3000\,\mathrm{\AA}$, the spectrum is slightly elevated, possibly due to Fe\,{\sc ii} complex \citep{Perez-Gonzalez2026}.

\begin{figure*}[ht!]
\centering
\includegraphics[width=2.1\columnwidth]{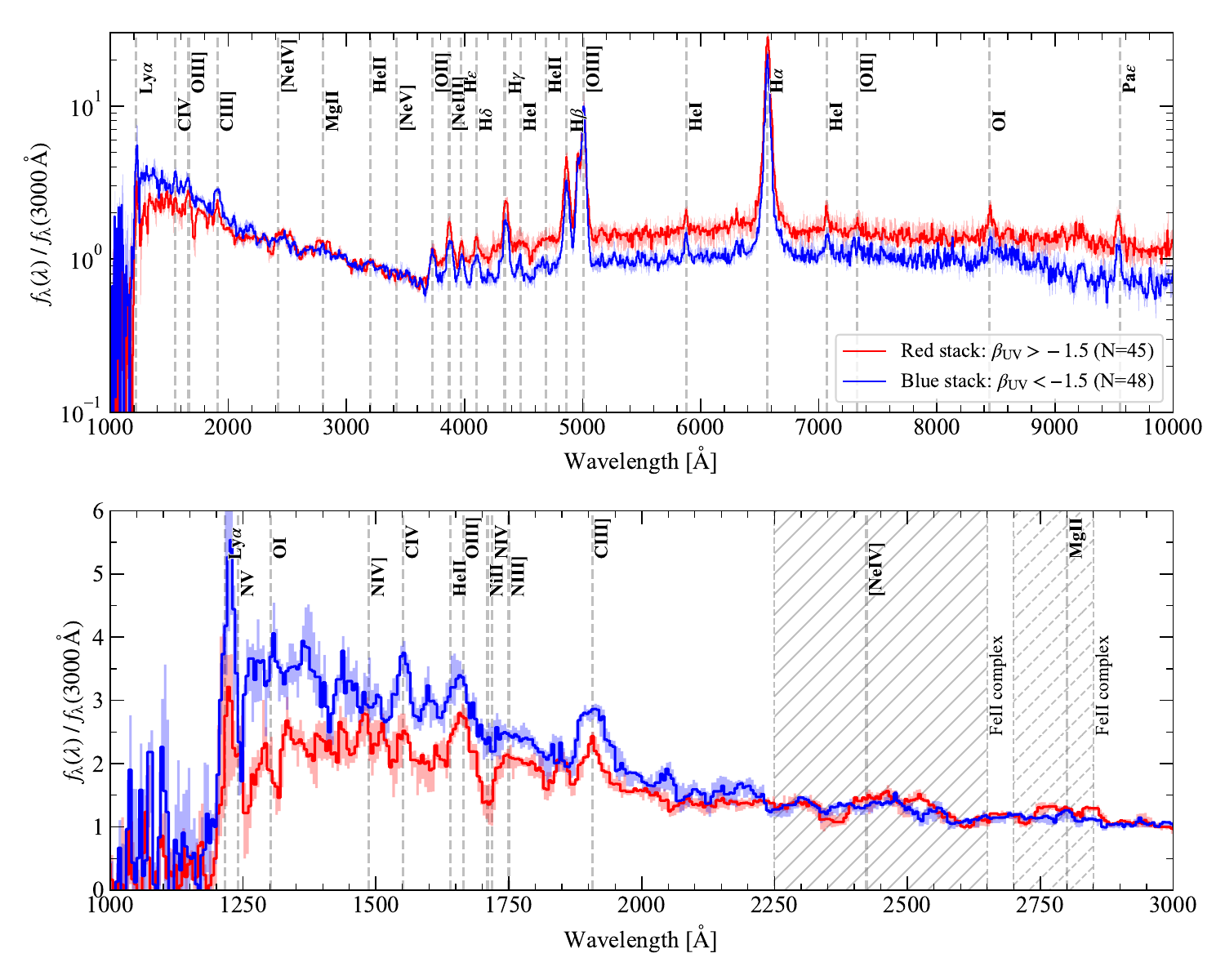}
\caption{Same as Figure~\ref{fig:stack_all}, but for the LRD subsamples divided by the UV slopes. The red and blue spectra are those for red ($\beta_\mathrm{UV}>-1.5$, N=45) and blue ($\beta_\mathrm{UV}<-1.5$, N=48) UV slopes, respectively.}
\label{fig:stack_beta}
\end{figure*}

\begin{figure*}[ht!]
\centering
\includegraphics[width=2.1\columnwidth]{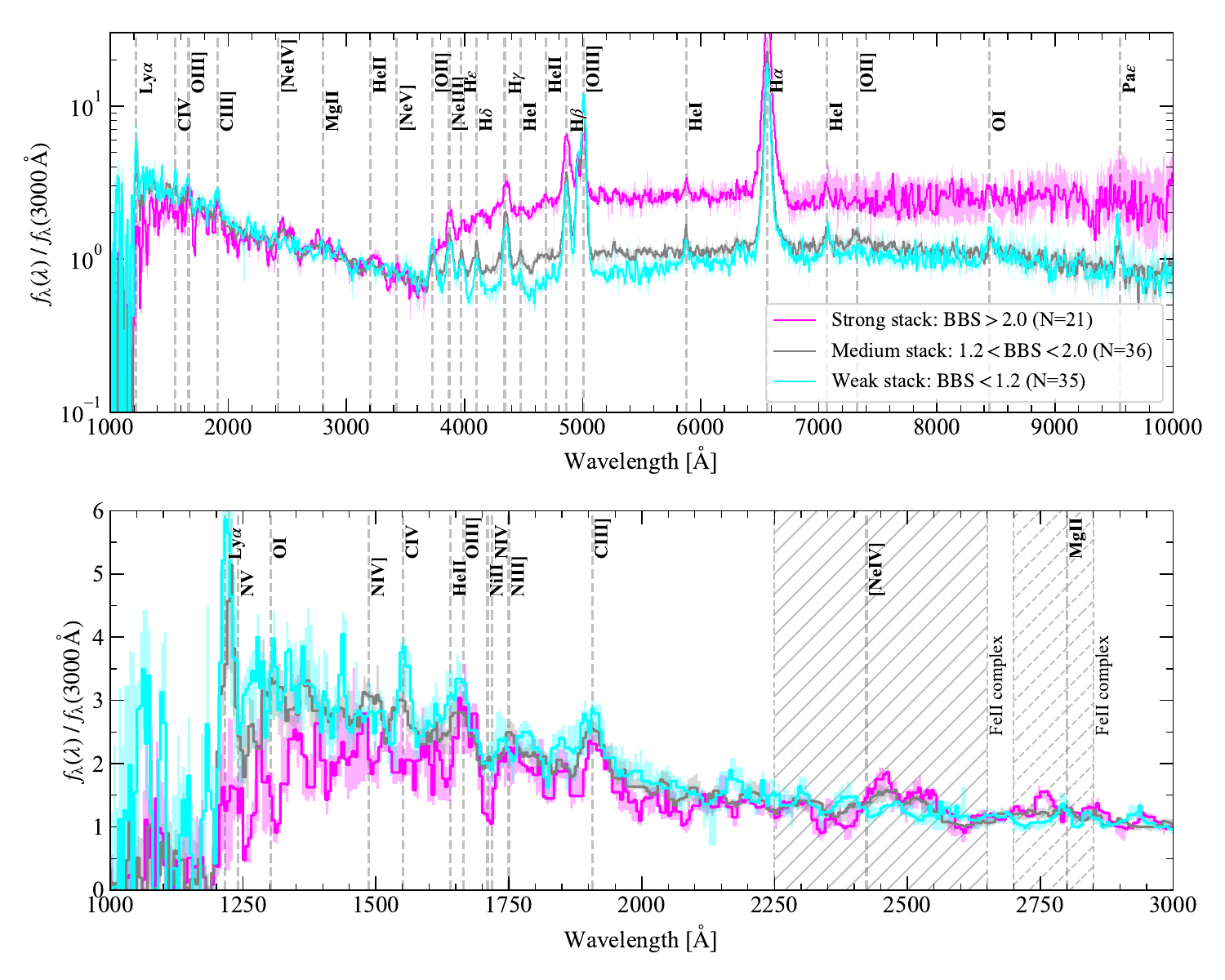}
\caption{Same as Figure~\ref{fig:stack_all} but for the subsamples divided by the Balmer break strength (BBS), which is defined as the flux ratio between $3620\text{--}3720\,\mathrm{\AA}$ and $4000\text{--}4100\,\mathrm{\AA}$. The magenta, gray, and cyan spectra are those of subsamples with the strong ($\mathrm{BBS}>2$, N=21), intermediate ($1.2<\mathrm{BBS}<2$, N=36), and weak ($\mathrm{BBS}<1.2$, N=35) Balmer breaks, respectively.}
\label{fig:stack_balmer}
\end{figure*}

Figure~\ref{fig:stack_beta} shows the stacked spectra for the red ($\beta_\mathrm{UV}>-1.5$, N=45, red) and blue ($\beta_\mathrm{UV}<-1.5$, N=48, blue) UV-slope subsamples, and Figure~\ref{fig:stack_balmer} displays those for the strong ($>2$, N=21, magenta), intermediate ($1.2\text{--}2$, N=36, gray), and weak ($<1.2$, N=35, cyan) Balmer-break subsamples. In the following subsections, we focus on the UV continuum and emission lines such as Ly$\alpha$, C\,{\sc iii}], C\,{\sc iv}, and Fe\,{\sc ii} complex.

\subsubsection{Continuum properties}
\label{sec:result_stack_cont}

% UV continuum and downturn
For continuum emission, we first estimate the UV slope of the stacked spectrum for all the LRDs. We derive $\beta_\mathrm{UV}=-1.35^{+0.23}_{-0.23}$, which is significantly redder than that of normal star-forming galaxies and consistent with those estimated for individual spectra. One interesting feature in the UV continuum is a downturn at the redward of Ly$\alpha$. This feature is similar to the Ly$\alpha$ damping wing, which is caused by a dense neutral gas along the line-of-sight, although Ly$\alpha$ itself is clearly detected. We note that this downturn signature is also detected when we limit the LRD sample to $z<6$, suggesting that it is not primarily due to intergalactic medium (IGM) absorption during the epoch of reionization. To quantify the depth of the downturn, we calculate Ly$\alpha$ damping wing parameter $D_\mathrm{Ly\alpha}$ introduced by \citet{Heintz2025}:
\begin{equation}
\label{eq:d_lya}
D_\mathrm{Ly\alpha}=\int^{1350\,\mathrm{\AA}}_{1180\,\mathrm{\AA}}{(1-f_\lambda/f_\mathrm{\lambda,cont}) d\lambda},
\end{equation}
where $f_\lambda$ and $f_\mathrm{\lambda,cont}$ are rest-frame observed and model continuum flux density, respectively. For the continuum, we adopt the power-law spectrum derived from the UV slope estimate of the stacked spectrum. Originally, $D_\mathrm{Ly\alpha}$ is proposed as an indicator of the damping wing absorption such that the large ($\gtrsim55\,\mathrm{\AA}$) or intermediate ($\sim35\text{--}55\,\mathrm{\AA}$) value suggests the existence of damped Ly${\alpha}$ systems (DLAs) or IGM absorption along the line-of-sight, while the small ($\sim0\text{--}35\,\mathrm{\AA}$) or negative value implies Ly$\alpha$ emission. We note, however, that $D_\mathrm{Ly\alpha}$ can also be large without damping wing absorption when the intrinsic continuum shape has a downturn around Ly$\alpha$, which is observed in nebular-dominated galaxies due to the two-photon process \citep{Cameron2024,Katz2025}. Therefore, we just treat this quantity as a measure of the depth of the downturn rather than damping wing absorption by DLA or IGM. We derive $D_\mathrm{Ly\alpha}=34.2^{+4.1}_{-4.1}\,\mathrm{\AA}$ for our stacked spectrum, indicating that the LRDs, on average, exhibit a significant downturn feature around Ly$\alpha$.

To further investigate how these spectral features correlate, we then focus on the LRD subsamples based on UV slope and Balmer break strength. As seen in Figure~\ref{fig:stack_beta}, the stacked spectrum with the redder UV-slope subsample shows a Balmer break stronger than that observed in the stacked spectrum based on the blue UV-slope subsample. Moreover, the downturn at the redward of Ly$\alpha$ looks deeper for the red UV slope subsample. These may indicate the physical connection between the optical and UV origins. Similar trends are exhibited for the Balmer-break subsamples: subsamples with stronger Balmer breaks show redder UV slopes and deeper downturns (Figure~\ref{fig:stack_balmer}).

\begin{figure}[ht!]
\centering
\includegraphics[width=\columnwidth]{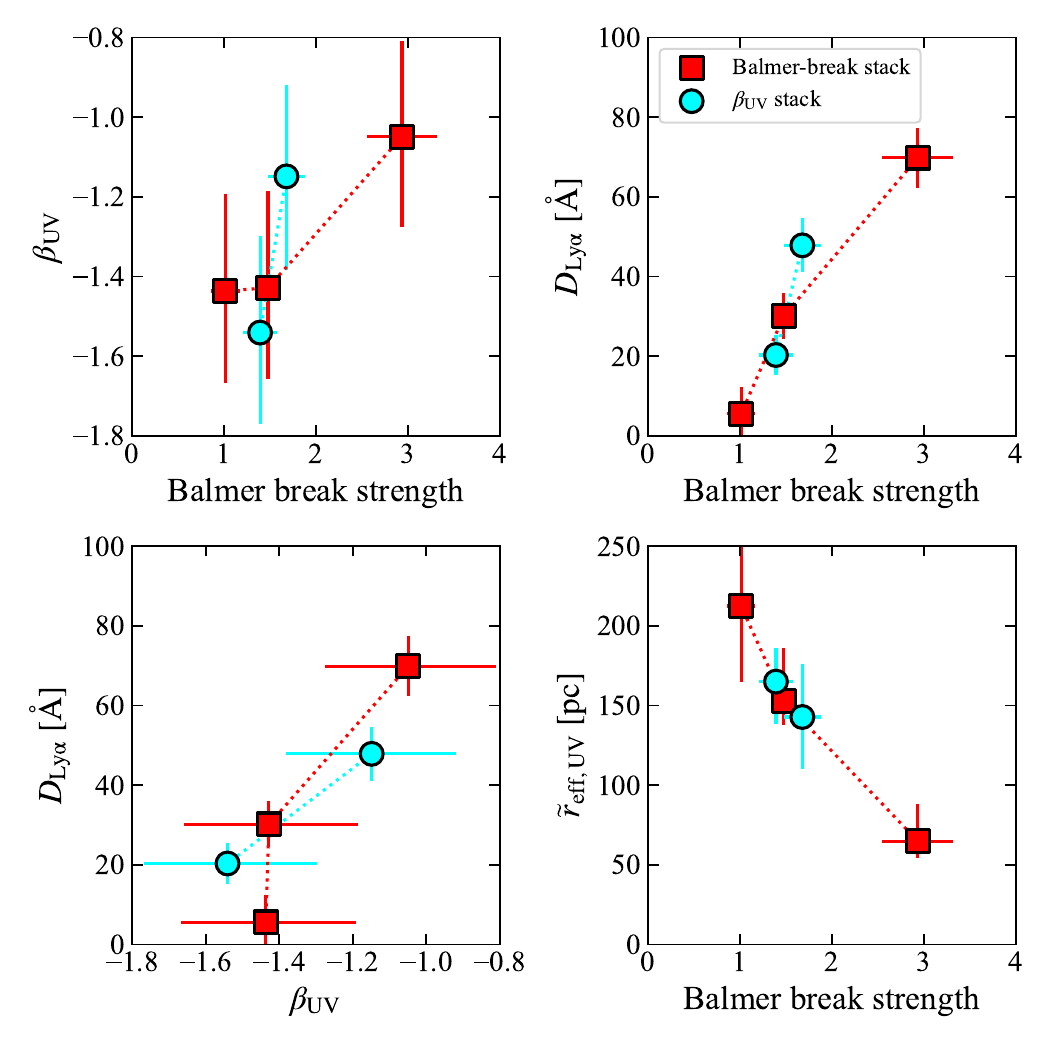}
\caption{Correlations among the Balmer break strength, UV slope ($\beta_\mathrm{UV}$), depth of downturn around Ly$\alpha$ ($D_\mathrm{Ly\alpha}$), and median UV size. The red squares and cyan circles represent the stacked spectra based on Balmer-break-strength and UV-slope subsamples, respectively. Clear correlations exist among these quantities: LRDs with stronger Balmer breaks have redder UV slopes, deeper downturns, and smaller sizes.}
\label{fig:stack_comp}
\end{figure}

To quantify these trends, we measure the UV slope, the Balmer break strength, and the damping wing parameter for the stacked spectra for each subsample. In Figure~\ref{fig:stack_comp}, we show the correlation between these three parameters.
As expected, these three parameters show a clear positive correlation. In addition, we also examine how the median UV sizes among the subsample depend on the Balmer break strength in the bottom right panel of Figure~\ref{fig:stack_comp}. We use LRDs detected in the UV images with $\mathrm{SN}>5$ to ensure reliable UV size measurements and a sufficient number of sources for statistical analysis. We find a clear trend that LRDs with the stronger Balmer break are more compact in the UV image.
If the Balmer break strength is an indicator of the dominance of the central source, this may support the idea that the red UV slope and deep downturn feature originate from the emission from the central compact source. We discuss this possibility in the Section~\ref{sec:discussion_model}.

\subsubsection{Ly$\alpha$ emission line}
\label{sec:result_stack_lya}
Stacked spectra of LRDs exhibit clear Ly$\alpha$ emission. To quantify the strength of this emission line, we measure the EW of the Ly$\alpha$. Since Ly$\alpha$ often has non-Gaussian profiles, instead of model fitting, we measure normalized Ly$\alpha$ flux by summing the fluxes of the continuum-subtracted spectra over $1200\text{--}1250\,\mathrm{\AA}$. As discussed in Section~\ref{sec:result_stack_cont}, the LRD stacked spectra exhibit the downturn at the redward of Ly$\alpha$. This may lead to an overestimation of the continuum if we estimate the continuum level at Ly$\alpha$ by extrapolating the power-law fit from the longer wavelength range (e.g., $1260-3500\,\mathrm{\AA}$). Instead, we estimate the local continuum level around Ly$\alpha$ from the following two methods: 1) 4th-order spline interpolation over $1150\text{--}1450\,\mathrm{\AA}$ removing the Ly$\alpha$ wavelength range, and 2) linear extrapolation from $1250\text{--}1350\,\mathrm{\AA}$. We note that the main source of uncertainty in the EW estimate is the continuum measurement at the PRISM resolution rather than the line SN.

\begin{figure}[ht!]
\centering
\includegraphics[width=1.0\columnwidth]{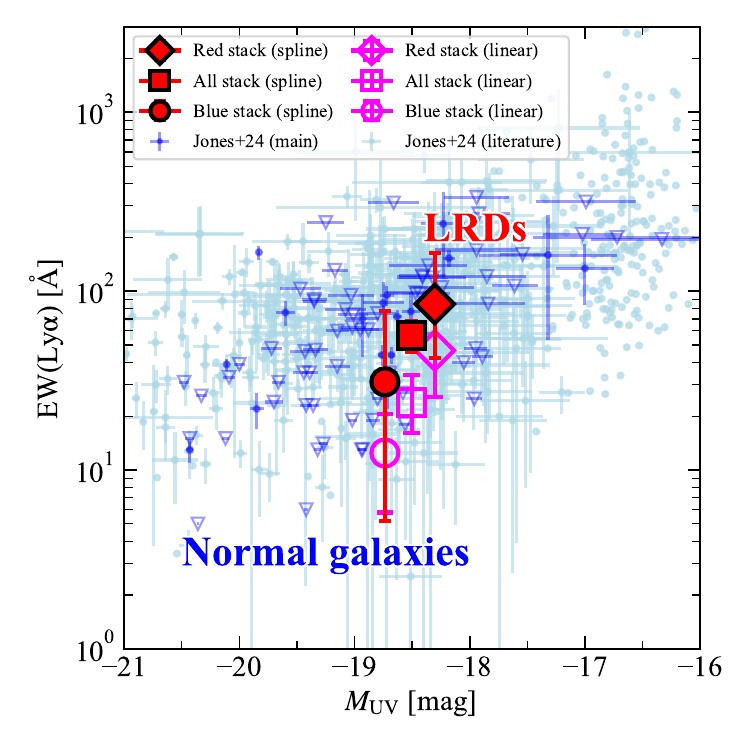}
\caption{Ly$\alpha$ EWs measured for UV-slope subsamples. The circle, square, and diamond represent the stacked spectra for the blue, all, and red subsamples, respectively. The red-filled and magenta-open symbols are the EWs calculated with the spline and linear continuum estimates, respectively. The median UV magnitudes are $\sim -18.5\,\mathrm{mag}$ for all three subsamples, but we slightly offset them for clarity. The lightblue dots are Ly$\alpha$ EWs estimated for star-forming galaxies at $z>4.5$ taken from \citet{Jones2024} and their references.}
\label{fig:lya}
\end{figure}

In Figure~\ref{fig:lya}, we show the measured Ly$\alpha$ EWs for the different UV-slope subsamples. The EWs of Ly$\alpha$ are $30\text{--}80\,\mathrm{\AA}$ and $10\text{--}50\,\mathrm{\AA}$ for spline and linear continuum estimates, respectively. While the stack for the all LRD sample has $\mathrm{EW(Ly\alpha})=56.5^{+11.6}_{-11.0}\,\mathrm{\AA}$ with spline continuum estimate, the red (blue) UV-slope subsample stack shows slightly higher (lower) EWs of $84.8^{+78.3}_{-42.3}\,\mathrm{\AA}$ ($31.1^{+46.0}_{-25.9}\,\mathrm{\AA}$), though comparable within uncertainties. We compare our $\mathrm{EW}(\mathrm{Ly\alpha})$ estimates to those measured for star-forming galaxies at $z>4.5$ (see \citealp{Jones2024} and their references). The LRD subsamples are located within the $\mathrm{EW}(\mathrm{Ly\alpha})$ distribution of star-forming galaxies, consistent with \citet{Asada2026}.

\begin{figure}[h!]
\centering
\includegraphics[width=\columnwidth]{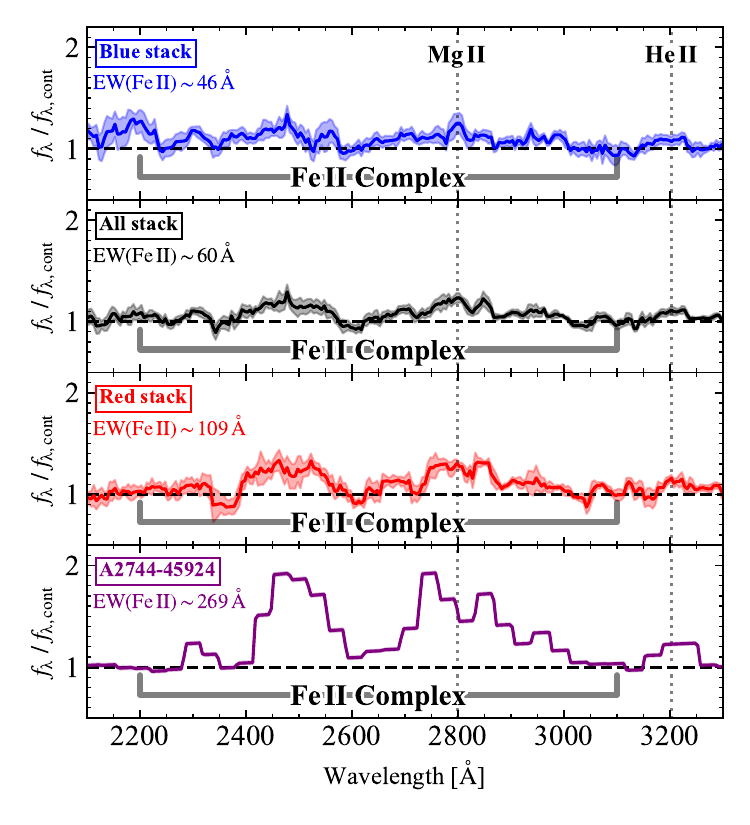}
\caption{Zoom-in view of the LRD spectra around the UV Fe\,{\sc ii} complex. Stacked Spectra for the blue, all, and red subsamples are shown from top to bottom. Additionally, the spectrum of A2744-45924 is also displayed. Spectra are normalized to a continuum level. The wavelengths of the UV Fe\,{\sc ii} complex, Mg\,{\sc ii}$\lambda2800$, and He\,{\sc ii}$\lambda3203$ are highlighted.} 
\label{fig:feii_spec_comp}
\end{figure}

\subsubsection{C\,{\sc iii}{\rm ]} and C\,{\sc iv} emission lines}
\label{sec:result_stack_ciii}
% Emission lines
The strengths of high-ionization carbon lines are indicators of active star formation and AGN signatures. We measure the EWs of C\,{\sc iii}] and C\,{\sc iv} emission lines observed in the stacked spectrum for all the LRDs, assuming their line shapes are single Gaussian. We find them to be $21.4\pm1.9\, \mathrm{\AA}$ and $8.1\pm2.3\, \mathrm{\AA}$, respectively, consistent with previous works \citep{Perez-Gonzalez2026,Sun2026}. According to an AGN diagnostic in \citet{Nakajima2018}, these relatively high EWs are comparable to the upper limits observed in star-forming galaxies (i.e., $20\,\mathrm{\AA}$ and $10\,\mathrm{\AA}$ for C\,{\sc iii}] and C\,{\sc iv}, respectively) and rather similar to the AGNs. The line ratio of $\mathrm{C}\,\text{\textsc{iv}}/\mathrm{C}\,\text{\textsc{iii}]}=0.50$ is consistent with both AGN and star-forming galaxy spectra \citep{Nakajima2018} but clearly smaller than that of quasars ($\gtrsim1$; \citealp{Onoue2020}). These carbon lines may not originate solely from either the AGN or the host galaxy, but instead may include contributions from both components. For the red (blue) UV-slope subsample, the results are similar: $\mathrm{EW(C\,\text{\textsc{iv}})}=8.0\pm3.1\, \mathrm{\AA}$ ($9.9\pm2.3\, \mathrm{\AA}$),  $\mathrm{EW(C\,\text{\textsc{iii}]})}=22.2\pm2.6\, \mathrm{\AA}$ ($27.9\pm2.9\, \mathrm{\AA}$), and $\mathrm{C}\,\text{\textsc{iv}}/\mathrm{C}\,\text{\textsc{iii}]}=0.46$ ($0.49$), respectively. For more conclusive diagnostics, we require He\,{\sc ii}$\lambda1640$ information, which is heavily blended with O\,{\sc iii]}$\lambda1665$ under the PRISM resolution.

\subsubsection{UV Fe\,{\sc ii} complex}
\label{sec:result_stack_feii}

The UV Fe\,{\sc ii} complex is typically observed in AGN spectra as pseudo-continuum, and detection of these lines strongly indicates the AGN-like emission (e.g., \citealp{Dong2011,Shen2014,Onoue2020}). We measure the strength of the UV Fe\,{\sc ii} complex at $2200\text{--}3100\,\mathrm{\AA}$. Since
Mg\,{\sc ii}$\lambda2800$, He\,{\sc ii}$\lambda3203$, and the UV Fe\,{\sc iii} complex are blended to the Fe\,{\sc ii} complex, we simultaneously fit these lines with a power-law continuum over $2000-3500\mathrm{\,\AA}$ using \texttt{fantasy} code \citep{Ilic2023}. For the Fe\,{\sc ii} flux, we integrate the best-fitted model spectrum over $2200-3100\mathrm{\,\AA}$.

\begin{figure*}[ht!]
\centering
\includegraphics[width=2.1\columnwidth]{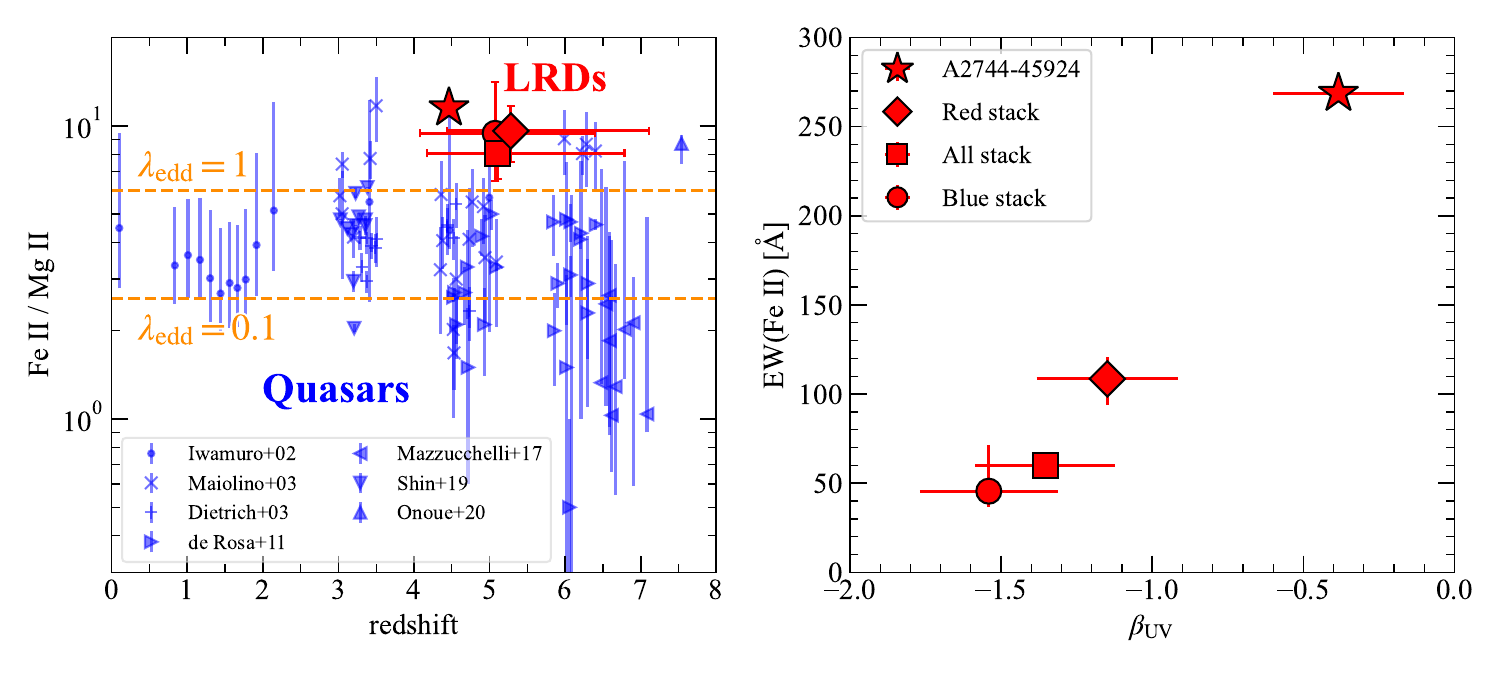}
\caption{\textit{Left panel:} measured Fe\,{\sc ii}/Mg\,{\sc ii} ratios for stacked spectra of LRDs. The circle, square, and diamond represent the stacked spectra for the blue, all, and red subsamples, respectively. Additionally, the flux ratio for A2744-45924 is also shown as the star. For stacked spectra, the median, 16th, and 84th redshift percentiles for each subsample are shown as the x-values. The orange dashed lines indicate the flux ratios expected from a scaling relation in \citet{Sameshima2017} under the Eddington ratios of $1$ and $0.1$. For comparison, Fe\,{\sc ii}/Mg\,{\sc ii} measurement for quasars in the literature are plotted as the blue symbols. LRDs are located on the upper side of the quasar distribution. \textit{Right panel:} EWs of Fe\,{\sc ii} against UV slope of each stacked spectrum. The EWs of Fe\,{\sc ii} increase with the UV slope.}
\label{fig:feii}
\end{figure*}

Figure~\ref{fig:feii_spec_comp} shows the zoom-in view of the spectra around the UV Fe\,{\sc ii} complex normalized by the continuum level for all, blue, and red UV-slope subsamples and A2744-45924, one of the brightest LRDs dubbed as `monster', \citep{Labbe2024}. The UV Fe\,{\sc ii} emissions are clearly detected, and their strengths vary among the displayed spectra. In the left panel of Figure~\ref{fig:feii}, we show measured Fe\,{\sc ii}/Mg\,{\sc ii} for all, blue, and red UV-slope subsamples. These stacked spectra show $\mathrm{Fe\,\text{{\textsc{ii}}}/Mg\,\text{{\textsc{ii}}}}\sim 8\text{--}10$ without any significant difference among the subsamples. We also examine the spectrum of A2744-45924 and find Fe\,{\sc ii}/Mg\,{\sc ii} to be $12$. If we use a scaling relation presented in \citet{Sameshima2017}, the expected Eddington ratios of three subsamples and A2744-45924 exceed unity, meaning super-Eddington accretion. If a scaling relation in \citet{Dong2011} is instead assumed, these flux ratios correspond to the Eddington ratio of $0.3\text{--}0.6$.

In the right panel of Figure~\ref{fig:feii}, we compare the EWs of Fe\,{\sc ii}. We find that LRDs with redder UV slopes have larger $\mathrm{EW(Fe\,\text{{\sc ii}})}$. These results may imply that BH activities are similar across different LRD subsamples, but the relative contributions of the host galaxy greatly differ. We discuss this point in the Section~\ref{sec:discussion_model}.

\section{Discussion}
\label{sec:discussion}

\subsection{Origin of UV continuum}
\label{sec:discussion_model}

In this study, we investigate the UV continuum shape, emission lines, and size of LRDs. We find that the LRDs have redder UV slopes ($\beta_\mathrm{UV}\sim-1.4$) and UV sizes (several hundred pc or less) smaller than normal star-forming galaxies. Their continuum emission shows the characteristic downturn shapes at the redward of Ly$\alpha$, similar to the Ly$\alpha$ damping wing or the nebular continuum with the two-photon process. Moreover, we detect Fe\,{\sc ii} complex, which is seen in AGN spectra, while C\,{\sc iv}, another representative emission line emitted from the AGN, is weak against C\,{\sc iii]}. We attempt to explain these observed UV features in a self-consistent scenario.

%LRD hosts are not dusty SFGs
The red UV slope and small size could be explained if the host is a compact and dusty star-forming galaxy. Assuming the dust attenuation of \citet{Calzetti2000} and an unattenuated UV slope of $\beta_\mathrm{UV, int}=-2.23$ \citep{Meurer1999}, we estimate $A_\mathrm{V}\sim 0.8\,\mathrm{mag}$ to explain the observed UV slope of $\beta_\mathrm{UV,obs}=-1.4$ (i.e., the all LRD stack) and $A_\mathrm{V}\sim 1.7\,\mathrm{mag}$ for an extreme case of $\beta_\mathrm{UV,obs}=-0.4$ (i.e., A2744-45924). This non-zero dust attenuation, however, is inconsistent with many expectations for the LRD host. \citet{Nikopoulos2025} have shown that the narrow Balmer line decrements of LRDs are comparable to those in the case B recombination, suggesting that the emission from the host is not strongly attenuated, although weak attenuation is still allowed within uncertainties. Under the assumption that the UV emission of LRDs largely originates in host star-forming galaxies, \citet{Sun2026} have shown that an LRD's host exhibits strong emission lines (e.g., $\mathrm{EW(C}\,\text{\textsc{iii}]}\lambda1908)\sim12\,\mathrm{\AA}$ and $\mathrm{EW([O}\,\text{\textsc{iii}}\mathrm{]}\lambda5007)\sim1100\,\mathrm{\AA}$) and thus is likely a young, star-forming dwarf galaxy with $\sim10^{8}\,M_{\odot}$, which is typically dust-poor.
The significant detection of Ly$\alpha$ in the stacked spectra also supports the dust-poor hosts. 
Moreover, if the red UV slopes originate from a dusty host galaxy, it is difficult to explain the observed correlations between UV slopes and various physical properties, such as Balmer break strength, depth of the downturn around Ly$\alpha$, effective radius, and iron emission line strength.
Based on these results, we argue that the red UV slope is not solely explained by the dust in the star-forming host galaxy, but that other intrinsically red emission is required.

Other than dust attenuation, nebular continuum from dense ionized gas (e.g., $n_\mathrm{e}\sim 10^{5}\,\mathrm{cm^{-3}}$) and/or leaked emission from AGN accretion disk can also produce red UV slopes (e.g., $-1.5 \lesssim \beta_\mathrm{UV} \lesssim 0$).
Based on these findings, we examine three possible origins that may explain the observed UV continuum: (A) dense ionized gas, (B) dense ionized gas \& AGN, and (C) nebular-dominated host galaxy. We show the schematic view of these models in the top panels of Figure~\ref{fig:model_comp}, and the details are as follows:

\begin{figure*}[htbp]
    \centering
    \includegraphics[width=2.1\columnwidth]{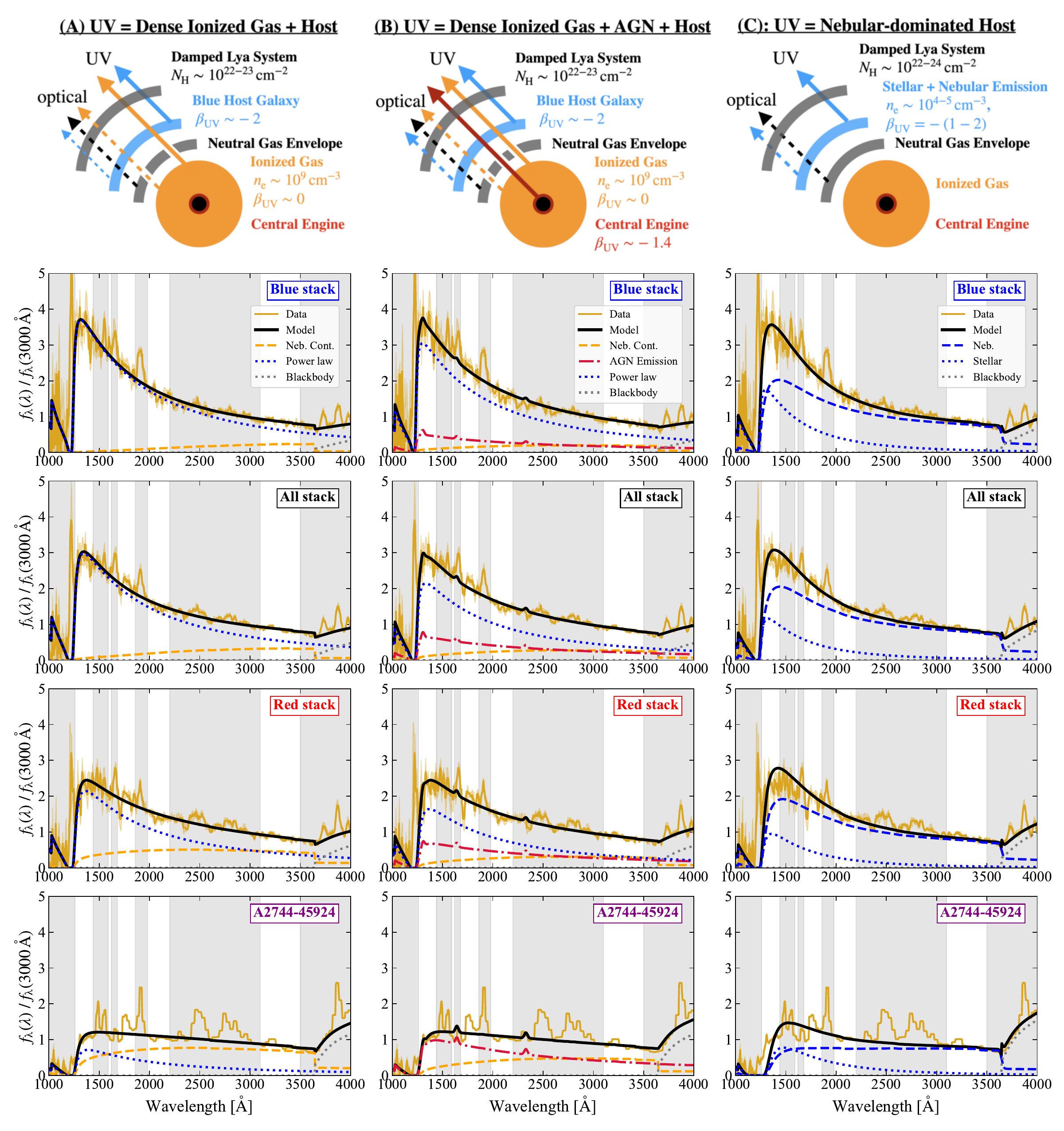}
    \caption{Spectral modeling of the UV continuum of LRDs. We show the three models from left to right: (A) Dense ionized gas \& host galaxy, (B) Dense ionized gas \& AGN \& host galaxy, and (C) Nebular-dominated host galaxy models. The top panels show a schematic illustration of each model. Below them, from top to bottom, stacked spectra for the blue ($\beta_\mathrm{UV}<-1.5$) subsample, the all sample, the red ($\beta_\mathrm{UV}>-1.5$) subsample, and the single spectrum of the A2744-45924 (dubbed as `monster' in the literature; \citealp{Labbe2024}) are shown as the yellow lines with uncertainties. The best-fitted models are shown in the black solid lines. The power laws with a slope of $\beta=-2$, corresponding to blue host galaxy spectra, are shown by blue dotted lines (models A-B). Nebular continuum from dense ionized gas (models A-B), and AGN emission (model B), are shown by the orange dashed lines and red dashed dotted lines, respectively. Incident stellar emission and the reprocessed nebular emission (model C) are shown by blue dotted and dashed lines, respectively. For clarity, emission lines are removed from the model C. The additional $5500\,\mathrm{K}$ blackbody spectrum (the gray dotted lines) is summed to each model, and the IGM absorption at $z=5$ \citep{Inoue2014} is also applied only for display purposes. Wavelength ranges masked during fitting are indicated by the gray shades.} 
    \label{fig:model_comp}
\end{figure*}

\begin{description}    
    \item[(A) Dense ionized gas \& host galaxy]\mbox{}

    The UV emission is attributed to the central dense ($\sim10^{9}\,\mathrm{cm^{-3}}$) ionized gas structure and the blue host galaxy. Due to escape channels (holes) or the clumpy nature of the dense neutral gas envelope, emission from the central ionized gas can escape and be observed. The nebular continuum from the dense ionized gas has an extremely red ($\beta_\mathrm{UV}\gtrsim0$) UV slope, and the diverse UV slopes are explained by relative contributions of the nebular continuum and the emission from the blue host. The downturn feature is explained by Ly$\alpha$ damping wing absorption of DLAs whose covering fraction is less than unity, which allows a part of the Ly$\alpha$ photon to escape.
    
    \item[(B) Dense ionized gas \& AGN \& host galaxy]\mbox{}
    
    In addition to the emission from the ionized gas in model A, photons from the central source are also considered. These photons can be observed due to the direct leakage or scattering (e.g., \citealp{Greene2024}). The central source is assumed to have an AGN-like spectrum, including emission from the accretion disk and surrounding nebula.

    \item[(C) Nebular-dominated host galaxy] \mbox{}
    
    The UV emission is attributed to the compact ($\lesssim 200\,\mathrm{pc}$) host galaxy, which is dominated by the nebular emission. The nebular continuum is emitted from the star-forming clumps with a high electron temperature ($T_\mathrm{e}\gtrsim10^{4}\,\mathrm{K}$) and a high electron density ($\sim10^{4-5}\,\mathrm{cm^{-3}}$) in the dust-poor host galaxy. These nebular conditions produce a red UV slope (e.g., $\beta_\mathrm{UV}\sim-1.5$) due to the Balmer continuum and a downturn shape at the redward of Ly$\alpha$ by boosting the two-photon process that has a bump at $1500\,\mathrm{\AA}$ in $f_\mathrm{\lambda}$ (\citealp{Cameron2024,Katz2025}). The emissions from the dense ionized gas and the central engine are subdominant, possibly due to the strong absorption by the dense neutral gas envelope.
\end{description}

We develop simple SED models describing these possibilities and test whether they fit the observed LRD spectra. We emphasize that the purpose of this modeling is not to develop exact physical models of LRDs but rather to examine whether our proposed models qualitatively reproduce the observed spectra. For model A, we assume the spectrum is a simple sum of the nebular continuum from the dense ionized gas and blue host emission. For simplicity, we assume the host emission follows a power law. The model A is expressed as:

\begin{equation}
    \label{eq:model_a}
    f_{\lambda}=f_{\lambda,\mathrm{neb}}\cdot \mathrm{NC}(T_\mathrm{e},n_\mathrm{e};\lambda)+f_{\lambda,\mathrm{power}} \left(\frac{\lambda}{1500\, \mathrm{\AA}} \right)^{-\beta},
\end{equation}
where $\mathrm{NC}$ is a nebula continuum model calculated by a python library \texttt{pyneb} \citep{Luridiana2015,Morisset2020,Mendoza2023} with the fixed electron density of $n_\mathrm{e}=10^{9}\,\mathrm{cm^{-3}}$ \citep{Naidu2025,X.Ji2025}.\footnote{The electron density of dense neutral gas envelope is suggested to be $n_\mathrm{e}\gtrsim10^{9}\,\mathrm{cm^{-3}}$ \citep{Naidu2025,X.Ji2025}. While we adopt $n_\mathrm{e}=10^{9}\,\mathrm{cm^{-3}}$, the results are almost unchanged if we instead assume a larger electron density (e.g., $n_\mathrm{e}=10^{10}\,\mathrm{cm^{-3}}$).} $f_{\lambda,\mathrm{neb}}$ and $f_{\lambda,\mathrm{power}}$ are normalizations for the two terms. We assume a power-law slope $\beta = -2$, which is typical for star-forming galaxies \citep{Bouwens2014}.
Additionally, we adopt the Ly$\alpha$ damping wing absorption modeled by \citet{Totani2006} with a free parameter of the hydrogen column density $N_\mathrm{H{\sc i}}$. This is needed to explain the downturn feature at the redward of Ly$\alpha$ despite the blue host emission.
For model B, we introduce AGN emission by adding a term $f_\mathrm{\lambda,AGN}\cdot \mathrm{AGN}(\lambda)$ to model A, where $f_\mathrm{\lambda,AGN}$ is a normalization. We adopt the SED model with $\beta_\mathrm{UV}\sim-1.4$ calculated by \citet{Inayoshi2022}, which represents the SED of super-Eddington accreting BHs with $\sim 10^6\,M_{\odot}$.
For model C, we adopt the model spectra of nebular-dominated galaxies presented in \citet{Katz2025}. The model spectra are computed by the \texttt{cloudy} code \citep{Chatzikos2023}, assuming the low-metallicity (i.e., 1\% solar) gas. Since high-temperature incident starlight is required to produce strong nebular emissions and consequently red UV slopes ($\beta_\mathrm{UV}>-2$), we adopt SED templates calculated with population III stellar models \citep{Larkin2023}. The templates are computed for different stellar temperatures ($T_\mathrm{stellar}\sim10^{4-5}\,\mathrm{K}$) and nebular electron density ($n_\mathrm{e}=10^{2-5}\,\mathrm{cm^{-3}}$). The Ly$\alpha$ damping wing absorption is applied.

For models A and B, we fit these SED models to the stacked spectra via an MCMC procedure, using the wavelength masks described in Section~\ref{sec:analysis_stack}. We construct best-fit models using the median of each parameter derived from the posterior distribution. For model C, we perform the least-squares fit. To investigate whether the models reproduce the spectral diversity of LRDs in terms of UV slope, we fit the models not only to the stacked spectrum of all LRDs but also to the stacked spectra of the red and blue UV-slope subsamples. In addition, we also fit the SED of A2744-45924, an optically bright LRD with a very red UV slope ($\beta_\mathrm{UV}\sim-0.4$) and small size ($r_\mathrm{eff,UV}<60\,\mathrm{pc}$), indicating the dominance of emission from its central component against the host.
For model B, we first fit the A2744-45924 spectrum without a power-law component. Then, when fitting other spectra, we fix the electron temperature and the ratio of $f_\mathrm{\lambda,AGN}/f_\mathrm{\lambda,neb}$ to the best-fitted values derived for A2744-45924. This is to solve the degeneracy between AGN and power-law components. 
In Figure~\ref{fig:model_comp}, we show the best-fit SEDs of models A, B, and C from left to right, with blue, all, red stacks, and the A2744-45924 spectra from top to bottom. 

Overall, models A and B successfully reproduce the observed continua with different UV slopes and downturn features at the redward of Ly$\alpha$. The expected electron temperature of dense ionized gas is $T_\mathrm{e}\sim 16,000\text{--}28,000\,\mathrm{K}$, and the downturn is reproduced by the damping wing with H\,{\sc i} column density of $\sim 10^{22.5}\,\mathrm{cm}^{-2}$. Models A and B also infer that redder UV slopes require a higher contribution from the emission of the central components (i.e., dense ionized gas and/or an AGN), relative to the blue host component.

For model C, the solutions with the high nebular electron density ($n_\mathrm{e}= 10^{4}\,\mathrm{cm^{-3}}$) and high stellar temperature ($T_\mathrm{stellar}\sim 80,000\text{--}100,000\,\mathrm{K}$) conditions are preferred. Similar to models A and B, this model also needs the Ly$\alpha$ damping wing absorption with $\sim 10^{22.5}\,\mathrm{cm}^{-2}$. For the A2744-45924 spectrum, even higher electron density ($n_\mathrm{e}\sim 10^{5}\,\mathrm{cm^{-3}}$) with high H\,{\sc i} column density $\sim 10^{24}\,\mathrm{cm}^{-2}$ solution is preferred. While the spectrum of the blue subsample is well reproduced by this model, the other spectra are not well fitted, especially below $1500\,\mathrm{\AA}$. This is due to the contribution of blue incident stellar light, which cannot be separated from redder nebular emission.

It is possible that modest additional dust attenuation ($A_{V}<0.5$) improves the fit of model C. However, the presence of such dust attenuation remains highly uncertain in low-metallicity galaxies (e.g., 1\% solar; \citealp{Cameron2024, Katz2025}) that exhibit dominant nebular emission \citep{Almeida2014, Zou2024}. Indeed, extremely metal-poor galaxies ($<10\%$ solar) host negligible dust, with $E(B-V) \sim 0.01\,\mathrm{mag}$ and thus $A_{V} \sim 0.04\,\mathrm{mag}$ \citep{Zou2024}. Another challenge of this model is the observed UV Fe\,{\sc ii} complex. While Fe\,{\sc ii} emission is often observed in quasar spectra, these lines are usually observed as absorption for normal galaxies (e.g., \citealp{Erb2012,Dong2011}). Moreover, it is unclear why the Balmer break strength and the UV continuum shape (e.g., $\beta_\mathrm{UV}$ and $D_\mathrm{Ly\alpha}$) are correlated, given that the UV and optical emission have distinct origins in this model. Therefore, it is likely difficult to explain the observed UV properties solely by host galaxy contributions.

\begin{figure*}[ht!]
\centering
\includegraphics[width=2.\columnwidth]{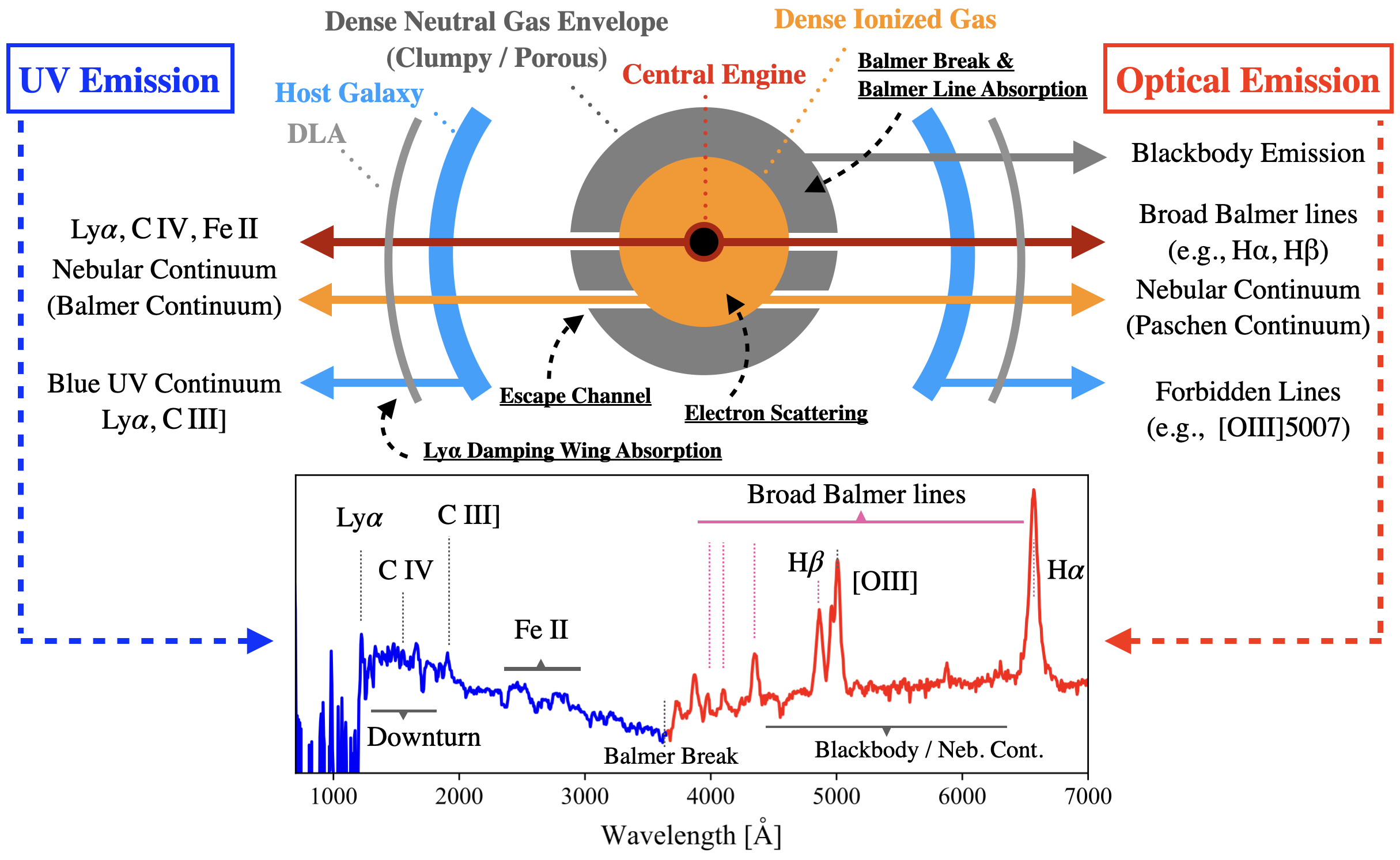}
\caption{Schematic illustration of origins of the rest-frame UV (left side) and optical (right side) emission from LRD based on a clumpy/porous BH envelope model (models A and B in Figure~\ref{fig:model_comp}) that explains UV spectral features investigated in this study and optical emission in the literature. Emissions from the central engine, dense ionized gas, dense neutral gas envelope, and host galaxy are accounted for. The stacked spectrum of the red UV-slope subsample is shown at the bottom for illustrative purposes.}
\label{fig:cartoon_lrd_emission}
\end{figure*}

\subsection{Implications for LRD structure}
\label{sec:implication}
As demonstrated in Section \ref{sec:discussion_model}, the leaked emission from the dense ionized gas and AGN (models A and B) can reproduce the UV continuum emission of LRDs. In Figure~\ref{fig:cartoon_lrd_emission}, we summarize the proposed origins of the emission in the rest-UV and rest-optical based on our findings and those discussed in the literature.

For the UV range, the Nebular (Balmer) continuum and the emission lines such as C\,{\sc iii}], C\,{\sc iv}, Fe\,{\sc ii}, which are produced by the central engine and the dense ionized gas, escape from the dense neutral gas envelope. A fraction of Ly$\alpha$ emission also originates from this leaked emission. From the host galaxies, bluer continuum as well as emission lines such as Ly$\alpha$ and C\,{\sc iii}] are emitted. The associated DLAs produce the downturn continuum shape around Ly$\alpha$.

For the optical range, the blackbody radiation from the pseudo-photosphere of the dense neutral gas envelope is observed. The Balmer break and the absorption features observed in the Balmer lines are produced around the neutral envelope. From the interior of the neutral gas envelope, the Balmer lines, which are broadened by the electron (Thomson) scattering in dense ionized gas, and the Paschen continuum, if it exists, would escape. In addition, some forbidden lines such as [O\,{\sc iii}]$\lambda5007$ are emitted by the host galaxy.

In the following sections, we discuss the consistency of the proposed models with other observational features and their implications for the physical picture of LRDs.

\subsubsection{high ionization emission lines}
Given that the central engine of an LRD is an AGN, emission lines with high ionized potential, such as C\,{\sc iv}, He\,{\sc ii}, [Ne\,{\sc iv}], and Fe\,{\sc iv}, are expected to be detected in LRDs' spectra if escape channels exist in the dense neutral gas envelope. While these lines are detected for a few LRDs (e.g., \citealp{Labbe2024, Tripodi2025, Torralba2026b, Tang2026}), they are absent in the majority of LRDs \citep{M.Tang2025, B.Wang2026}. One possible explanation is that these emission lines are diluted by the host galaxy emission. The best-fit solutions for models A and B suggest that the host galaxy component dominates the UV range, especially for bluer LRDs. The spectrum of A2744-45924, which exhibits high-ionization lines, shows a smaller contribution from the host than the other spectra.

Our observations of carbon lines are consistent with this explanation. We measure C\,{\sc iii}] and C\,{\sc iv} and find that the EWs of these lines do not clearly exceed the star-forming galaxy regime. The continuum emission from the star-forming host galaxy may decrease the EWs of these lines. We also find that the line ratios of C\,{\sc iv}/C\,{\sc iii}] are much smaller than those of the quasar populations, indicating the contamination of C\,{\sc iii}] emission from the host. Another possibility is that the central engine of an LRD is in a low-metal environment, resulting in intrinsically weak metal lines.

\subsubsection{Ly$\alpha$ emission and absorption}

Our models suggest that the DLAs with $N_\mathrm{H\,\text{\sc i}}\sim10^{22.5}\, \mathrm{cm^{-2}}$ are always associated with LRDs. The Ly$\alpha$ damping wing absorption is introduced into our models to explain the UV downturn around Ly$\alpha$. Since this downturn is detected in the median stacked spectra, DLAs likely cover most (or at least more than half) of the sightlines; otherwise, the downturn would be averaged out. Given such a high frequency, these DLAs are not merely projected objects along the sightline on an IGM scale but are physically associated with the LRDs on a galaxy scale. This suggests that LRDs are inherently gas-rich systems.

One important note, however, is that Ly$\alpha$ emission is clearly observed despite the presence of DLAs. If LRDs are dust-poor systems, Ly$\alpha$ photons may escape after resonant scattering, and the extended Ly$\alpha$ could be observed. \citet{Torralba2026a} have reported the detection of a weak Ly$\alpha$ halo with the narrow line profile around A2744-45924. Due to the spatial offset between Ly$\alpha$ and N\,{\sc iv} emissions, this Ly$\alpha$ halo has been suggested to originate from the host galaxy.
On the other hand, \citet{Tang2026} have reported that the line profile of Ly$\alpha$ emission detected from Abell2744-QSO1, an LRD at $z=7.04$, is likely broadened by the electron scattering rather than the resonant scattering and argued that Ly$\alpha$ emission from this object may escape from a dense neutral gas envelope. These findings suggest the diversity in the origins of Ly$\alpha$ emission and absorption from LRDs. To further test the coexistence of Ly$\alpha$ emission and DLAs, future deep, spatially resolved far-UV spectroscopy of individual LRDs is essential.

\subsubsection{UV size}

As shown in Section~\ref{sec:result_size}, the LRDs have a small UV size. If the leaked emission dominates the UV emission, the UV compactness of LRDs is naturally explained. Indeed, as shown in Figure~\ref{fig:stack_comp}, the LRDs with redder UV slopes (stronger Balmer breaks) are more compact, supporting this scenario if the UV slope (Balmer break strength) indicates a dominance of the central source. If a sufficient amount of ionizing photons escapes from the neutral gas envelope, those photons may ionize the gas at the outer layer of the dense neutral gas envelope. In that case, the additional nebular continuum would be produced, contributing to even redder UV slopes and compact UV morphology.

\subsubsection{Fe\,{\sc ii} emission}

In Section~\ref{sec:result_stack_feii}, we demonstrate that the line ratio of Fe\,{\sc ii}/Mg\,{\sc ii} is almost constant among LRDs, while the EWs of UV Fe\,{\sc ii} emission clearly increase with UV slopes. This may suggest that the UV emission from the central engine is similar across different LRDs, but the relative contributions of host galaxy emission make the UV properties of LRDs diverse. This is consistent with the results of our model analysis: redder UV slopes require higher contributions from emission from central components (i.e., ionized gas + central engine) relative to the host galaxy. Another possible interpretation of strong iron emission is that LRDs are iron-rich systems due to enhanced iron production in pair-instability supernovae, which occur in low-metallicity environments \citep{Langer2007, Hiramatsu2026}. Investigating the properties of the host galaxy may help us understand the origins of strong iron lines.

\subsubsection{X-ray emission}
LRDs are generally undetected in X-rays even by the deep Chandra observation \citep{Ananna2024,Akins2025,Sacchi2025,Kokubo2025,Maiolino2025}. Since X-rays, especially hard ones, penetrate neutral gas much more easily than UV photons (e.g., \citealp{Ueda2014}), one might expect X-ray photons to penetrate the dense neutral gas envelope if any escape channels are present, leading to Chandra detections. This discrepancy can be alleviated in several ways. One possibility is that the X-ray photons emitted by the central engine may be weakened due to Compton scattering within the dense ionized gas (e.g., \citealp{Rusakov2026}). Another idea is that the intrinsic X-ray emission is even fainter than expected in the previous works. While the bolometric luminosity has often been estimated using AGN templates (e.g., \citealp{Akins2025}), a blackbody-like optical SED would yield lower bolometric and corresponding X-ray luminosities \citep{Greene2026a}, reducing the tension. Furthermore, as discussed in Section~\ref{sec:result_stack_feii}, the Fe\,{\sc ii}/Mg\,{\sc ii} line ratios suggest the super-Eddington accretion. Under this regime, the hard X-ray emission would be suppressed (e.g., \citealp{Madau2024,Inayoshi2025b,Lambrides2026}), thereby explaining the lack of X-ray detections despite the presence of escape channels.

\subsubsection{Short-term variability}

The majority of LRDs lack the short-term (e.g., $\sim10^{2}$ days) variability \citep{Kokubo2025,Z.Zhang2025a,Liu2026}. If central engines are in the super-Eddington accretion state, the variability can be intrinsically small \citep{Inayoshi2025}. The emission from the host galaxy further dilutes the variability amplitude \citep{Kokubo2025}, leading to no clear detection of variability. In addition, due to the high electron density in the dense ionized gas, the photon diffusion time would be long, smoothing intrinsic variability to longer than a year \citep{Sneppen2026a}.

\subsubsection{Balmer/Paschen jump}

Our model suggests that the nebular continuum from the dense ionized gas contributes to the rest-UV emission of LRDs. One may expect the Balmer jump (i.e., a sharp discontinuity of the continuum at the Balmer limit) to be detected in the observed spectra (e.g., \citealp{Katz2025}). Indeed, \citet{Lin2026} have reported a possible detection of such a feature for LRDs at the local Universe ($z\lesssim0.2$). As shown in Figure~\ref{fig:model_comp}, while the nebular continua (orange dashed lines) show the Balmer jump, the host galaxy emission (blue dotted lines) and the blackbody-like emission from the pseudo-photosphere of the dense neutral gas envelope (gray dotted lines) may dilute this feature. On the other hand, \citep{Sneppen2026c} have argued that the optical continua of LRDs exhibit the Paschen jump, suggesting the nebular continua dominate the optical emission rather than the blackbody radiation. This is consistent with our expectation of escaped nebular continuum in the UV.

\section{Summary and Conclusion}

In this paper, we have examined the properties of rest-frame UV emission (i.e., continuum shapes, emission line strengths, and morphology) of $\sim100$ LRDs at $3.5<z<9.3$ and $-20<M_\mathrm{UV}<-15.5$, originally selected by \citet{de_Graaff2025c}. What we have found is as follows.

\begin{enumerate}
    \item We measure the UV slopes of LRD spectra at $1340\text{--}2700\,\mathrm{\AA}$ assuming a power-law form using JWST NIRSpec/PRISM spectra. We find that the UV slopes of LRDs are $\beta_\mathrm{UV}\sim -1.4$, which are systematically redder than normal star-forming galaxies by $\Delta \beta_\mathrm{UV}\sim 0.6$ with controlled redshifts and UV absolute magnitudes.
    
    \item We measure the UV size of LRDs assuming a 2D S\'ersic profile using JWST/NIRCam images. To focus on the UV emission from the main body of LRDs, which significantly contributes to their UV spectra, we carefully remove UV emission from possible companions by simultaneously fitting with multiple S\'ersic models. We find that the UV sizes of the LRDs are on average, one-fourth of star-forming galaxies with the same redshift and UV magnitudes. 

    \item We stack the LRD spectra and find that LRDs with stronger Balmer breaks have redder UV slopes, deeper downturns around Ly$\alpha$, and smaller sizes, suggesting that the Balmer break strength indicates the dominance of the emission from the central sources in the UV spectra.

    \item We detect UV emission lines from the stacked spectra, including Ly$\alpha$, C\,{\sc iii}], C\,{\sc iv}, Fe\,{\sc ii}, Mg\,{\sc ii}. While the EWs of Ly$\alpha$ are consistent with those of star-forming galaxies, those of C\,{\sc iii}] ($\sim20\,\mathrm{\AA}$) and C\,{\sc iv} ($\sim8\,{\mathrm{\AA}}$) are comparable to the upper limit that can be produced by star-forming galaxies. The line ratios of Fe\,{\sc ii}/Mg\,{\sc ii}$\sim8\text{--}10$ are significantly high compared to those of other AGN populations, suggesting that the emission from the central AGN significantly contributes to the UV spectra of LRDs.

    \item We propose three distinct origins of the red and compact UV emission of the LRDs: (A) the dense ionized gas with a blue host, (B) the AGN emission added to model A,  and (C) a nebular-dominated compact host galaxy. We construct simple continuum models, including the nebular continuum from the dense ionized gas, AGN emission, and the host galaxy emission.
    While Model C fails to reproduce the very red UV slopes observed in some LRDs due to a significant contribution from blue stellar emission, models A and B successfully reproduce the observed stacked spectra for subsamples with different UV slopes. We find that the diversity in the UV slopes is explained by different contributions from the blue host component, with decreasing contributions at redder UV slopes. The success of models A and B suggests that the emission from a dense ionized gas and/or an AGN leaks through the clumpy/porous neutral gas envelope. \\
\end{enumerate}

We shed light on the LRD structure using statistical samples of NIRSpec/PRISM spectra in the rest-frame UV range. 
Our results highlight the importance of carefully accounting for the UV emission from the central component of LRDs when investigating host-galaxy properties (e.g., stellar mass, star-formation history, etc.).
To further investigate the rest-UV properties of LRDs, we need higher resolution grating spectra. For example, resolving the narrow and broad components of permitted lines (e.g., C\,{\sc iv}) is important to distinguish models A and B. He\,{\sc ii}$\lambda1640$ can be used as a signature of AGN, which is significantly blended with the O\,{\sc iii]}$\lambda\lambda1661,1666$ doublet with NIRSpec/PRISM spectral resolution. Profiles of emission and absorption lines may provide detailed kinematics of the gas and envelope geometry. Moreover, deeper rest-frame UV observations enable us to examine the spectral features of individual LRDs, potentially revealing their diverse evolutionary stages or subclasses.

The exploration of the UV emission of LRDs provides insights into understanding properties of the host galaxies, the structure of the central envelope, and the conditions under which LRDs form. With the growing number of LRD samples and deep observations, the origins of LRDs will be better understood.

%% Please use the acknowledgment and contribution environments. This will 
%% be anonomyized when the "anonymous" style option is used. 
\begin{acknowledgments}

We thank Anna de Graaff et al. for the public release of the LRD catalog, and Hiroya Umeda, Hiroto Yanagisawa, Kazuhiro Shimasaku, Kei Ito, Suin Matsui, and Yuki Shibanuma for useful comments and discussions. 

This work is based on observations made with the NASA/ESA/CSA James Webb Space Telescope. The data were obtained from the Mikulski Archive for Space Telescopes at the Space Telescope Science Institute, which is operated by the Association of Universities for Research in Astronomy, Inc., under NASA contract NAS 5-03127 for JWST. These observations are associated with programs 1180, 1181, 1208, 1212, 1213, 1215, 1286, 1345, 1433, 2198, 2561, 2750, 2767, 4106, 4233, 5105, 5224, 6368, and 6585. The authors acknowledge these observing teams for developing their programs

The data products presented herein were retrieved from the Dawn JWST Archive (DJA). DJA is an initiative of the Cosmic Dawn Center (DAWN), which is funded by the Danish National Research Foundation under grant DNRF140.

MA and YH acknowledge support from the Japan Society for the Promotion of Science (JSPS) Grant-in-Aid for Scientific Research (24H00245), the JSPS International Leading Research (22K21349), the Sumitomo Foundation, the Ito Science Promotion Society, and the Yamaguchi Scholarship Foundation.
KI acknowledges support from the National Natural Science Foundation of China (12573015, W2532003), the Beijing Natural Science Foundation (IS25003), and the China Manned Space Program (CMS-CSST-2025-A09).
TST is supported by JSPS KAKENHI Grant Number JP25KJ0750 and the Forefront Physics and Mathematics Program to Drive Transformation (FoPM), a World-leading Innovative Graduate Study (WINGS) Program at the University of Tokyo.
\end{acknowledgments}

%\facilities{HST(STIS), Swift(XRT and UVOT), AAVSO, CTIO:1.3m, CTIO:1.5m, CXO}
\facilities{JWST(NIRSpec and NIRCam)}

\software{
    numpy \citep{numpy:2011},
    scipy \citep{scipy:2001},
    %pandas \citep{pandas:2010},
    %matplotlib \citep{matplotlib:2007}
    astropy \citep{astropy:2013,astropy:2018},
    msaexp \citep{msaexp},
    grizli \citep{grizli},
    galight \citep{Ding2021},
    pyneb \citep{Luridiana2015}.
          }


%\appendix

%\section{Appendix information}

%% For this sample we use BibTeX plus aasjournalv7.bst to generate the

%\bibliography{sample701}{}
\bibliography{bibtex_26}{}
\bibliographystyle{aasjournalv7}



\end{document}